\documentclass[reprint,aip,graphicx,cha,showpacs]{revtex4-1}

\usepackage{amsmath,amsfonts,bm,bbm}

\usepackage{float}
\usepackage{epsfig}
\usepackage{esint}
\usepackage{color}

\usepackage{multirow}

  \usepackage{paralist}
  \usepackage{epstopdf}
  \usepackage{graphics} 
 \usepackage[colorlinks=true]{hyperref}
 \hypersetup{urlcolor=blue, citecolor=red}
 \usepackage[latin1]{inputenc}
\usepackage[caption=false]{subfig}

 \usepackage{tikz-cd}

\newcommand{\R}{{\mathbb R}}
\newcommand{\calF}{{\mathcal F}}

\newcommand{\e}{{\rm e}}

\begin{document}
\title[Slow--fast  dynamics of a stochastic McKean model]{Slow--fast  dynamics of the McKean model with stochastic resetting and diffusion}

\author{Jude Swaby and Paul C. Bressloff}
\address{Department of Mathematics, Imperial College London, London SW7 2AZ, UK.}
\email{p.bressloff@imperial.ac.uk}

\date{\today}

% Abstract
\begin{abstract}
In this paper we investigate the combined effects of stochastic resetting and diffusion on a slow--fast dynamical system given by the piecewise-linear McKean model. That is, the fast variable $v$ is subject to Gaussian white noise with effective diffusivity $D$ and is reset to a fixed value $v_r$ at a random sequence of times generated from a Poisson process with rate $r$. Assuming that resetting occurs on the fast timescale, we freeze the slow variable \(w\) under an adiabatic approximation and determine the resulting non-equilibrium stationary state (NESS) of the fast variable as the solution of a modified Fokker--Planck equation. We show that the NESS can be expressed in terms of parabolic cylinder functions, whose asymptotic behavior allows us to recover the corresponding NESS without diffusion in the small-diffusion limit. The NESS is used to derive an averaged equation for the slow dynamics, whose solution converges to a stable fixed point \(w^*\) that depends on \(D\), \(r\) and \(v_r\). This fixed point effectively determines the long-time behaviour of the full system. Finally, we analyze the regime in which resetting occurs on the same timescale as the slow variable. We show how the slow variable now undergoes noisy oscillations due to resetting-induced switching between branches of the fast nullcline and derive the corresponding NESS for $w$ in the non-diffusive case.
\end{abstract}
 
\maketitle

\noindent {\bf In recent years there has been considerable interest in understanding stochastic resetting as a mechanism for maintaining a stochastic process in a nonequilibrium stationary state (NESS). The basic idea is to reset the underlying system of interest to a prescribed state at a random sequence of times that is typically generated by a homogeneous Poisson process with rate $r$. Most studies of resetting focus on systems consisting of one or more particles diffusing in some fixed domain or potential energy landscape. There has been relatively little work applying stochastic resetting within the context of dynamical systems theory. In this paper we consider a stochastic version of the piecewise linear McKean model, which is a classical example of a planar slow-fast system. The fast variable is subject to both stochastic resetting and Gaussian white noise fluctuations. Assuming that resetting occurs on the fast timescale, we use an adiabatic approximation to derive an averaged equation for the slow dynamics that depends on the mean of the fast variable with respect to a corresponding NESS in which the slow variable is fixed. The solution of the averaged equation thus converges to a stable fixed point that determines the long-time behaviour of the full system. Moreover, such behavior persists when the piecewise linear function of the McKean model is replaced by the cubic nonlinearity  of the FitzHugh-Nagumo model. Finally, we derive an NESS for the slow variable in the case of slow resetting and zero Gaussian noise. Assuming that the underlying deterministic system is oscillatory,
we show that the slow variable undergoes noisy oscillations due to resetting-induced switching between branches of the fast nullcline.}

\section{Introduction}

A topic of considerable current interest in nonequilibrium statistical physics is stochastic resetting, a mechanism whereby a system is reset to a prescribed state at a random sequence of times that is typically generated by a Poisson process at a constant rate $r$. The theory was first developed within the context of a Brownian particle instantaneously resetting to its initial position \cite{Evans11a,Evans11b,Evans14}. One major consequence of resetting is that the corresponding single-particle probability density converges to a nonequilibrium stationary state (NESS) that supports nonzero probability currents. The existence of a nontrivial NESS has subsequently been shown for a wide range of stochastic processes with resetting, see the review Ref. \onlinecite{Evans20}. Examples include non-diffusive processes such as Levy flights \cite{Kus14} and run-and-tumble processes \cite{Evans18,Bressloff20}, switching diffusions \cite{Bressloff20a}, non-Poissonian resetting protocols \cite{Eule16,Pal16,Nagar16}, and diffusion in a potential \cite{Pal15}. Note, however, that most of these studies are restricted to one-dimensional (1D) noisy dynamical systems.

An important subclass of two-dimensional (2D) dynamical systems is that of a slow--fast system, in which one variable evolves on a fast timescale and the other evolves on a slow timescale \cite{Wech20,Inahama25}. One classical example is given by the FitzHugh--Nagumo (FN) equations \cite{Fitz61,Nagumo62,Lacasa24}, which are a simplified version of the four-dimensional Hodgkin-Huxley equations \cite{HodgkinHuxley1952}. These so-called conductance-based systems are used to model the rapid voltage spike or action potential induced by the flow of ions across the cell membrane of a neuron \cite{KeenerSneyd2009,Terman}. Recently, one of us analyzed the deterministic FN equations supplemented by stochastic resetting in the fast voltage variable \cite{Bressloff25}. In particular, assuming that the resetting rate $r$ was the same order of magnitude as the relaxation rate of the fast dynamics, we showed that the corresponding slow dynamics reduced to an averaged equation that depended on the mean of the fast variable with respect to a corresponding 1D NESS. The solution of the averaged equation thus converged to an $r$-dependent fixed point. On the other hand, in the slow resetting regime, we found numerically that each resetting event triggered a trajectory that converged to an attractor of the underlying deterministic system. In the case of the FN model, the attractor was either a stable fixed point (excitable regime) or a stable limit cycle (oscillatory regime). Resetting triggered a sequence of pulses or action potentials in the excitable regime and fluctuations of the limit cycle in the oscillatory regime.

In contrast to most studies of stochastic resetting \cite{Evans20}, our analysis of the FN equations ignored the effects of Gaussian white noise fluctuations. In other words, the underlying dynamical system without resetting was taken to be deterministic rather than stochastic. Inclusion of both Gaussian white noise and resetting leads to a second-order Fokker--Planck (FP) equation for the fast variable in the adiabatic limit, whose NESS cannot be solved analytically due to the presence of the cubic nonlinearity. (Historically speaking, stochastic versions of the FN equations driven by Gaussian white noise in either the slow or fast variable and no resetting play an important role in understanding the effects of noise in excitable systems, see for example Refs. \onlinecite{Tuckwell98,Linder04}.) Therefore, in this paper, we replace the cubic nonlinearity by a piecewise linear function, resulting in the McKean model \cite{Tonnelier2002}. This allows us to analytically determine the NESS of the fast variable under the combined effects of stochastic resetting and Gaussian white noise. (In this paper we refer to the latter as a form of diffusion.) The slow dynamics can then be determined by averaging the fast variable with respect to the NESS along analogous lines to Ref. \onlinecite{Bressloff25}.

The structure of the paper is as follows. The stochastic McKean model with fast resetting is introduced in Sect. II. An exact expression for the NESS for the fast variable without diffusion is derived in Sect. III, which is then used to determine the slow dynamics via averaging. We thus establish that the slow variable converges to a stable fixed point as previously found for the FN model \cite{Bressloff25}. The analysis is then extended to the diffusive case in Sect. IV. We solve for the NESS in terms of parabolic cylinder functions, and again show how the corresponding averaged equation for the slow variable converges to a fixed point that now depends on both the diffusivity and resetting rate. In Sect. V  we use properties of parabolic cylinder functions to conduct a weak-diffusion asymptotic analysis of the NESS that recovers the non-diffusive case in the zero diffusion limit. Finally, in Sect. VI we analyze a case where resetting occurs on the same timescale as the evolution of the slow variable. Assuming that the McKean model operates in the oscillatory regime without diffusion, we show that the slow variable undergoes noisy oscillations due to resetting-induced switching between branches of the fast nullcline. We develop a probabilistic formulation of the switching dynamics and determine the resulting NESS for the slow variable.

\section{The stochastic McKean model with resetting}

\subsection{Deterministic model} Consider a deterministic 2D slow-fast dynamical system given in the (non-dimensionalized) form
\begin{subequations}
 \label{eqvw}
\begin{align}
\frac{dv}{dt} &= f(v) - w + I_0,\\
\frac{dw}{dt} &= \varepsilon\,(v - \gamma w), \label{eq:fn_w}
\end{align}
\end{subequations}
where $I_0$ is a constant applied input, $0<\varepsilon\ll 1$ sets the timescale separation, and $\gamma>0$ determines the slope of the linear $w$-nullcline. Within the context of neurodynamics, the fast variable $v$ represents a space-clamped membrane voltage, whereas the slow variable represents a slow recovery variable (often associated with a slow potassium current). 

\begin{figure}[b!]
    \centering
    \includegraphics[width=\linewidth]{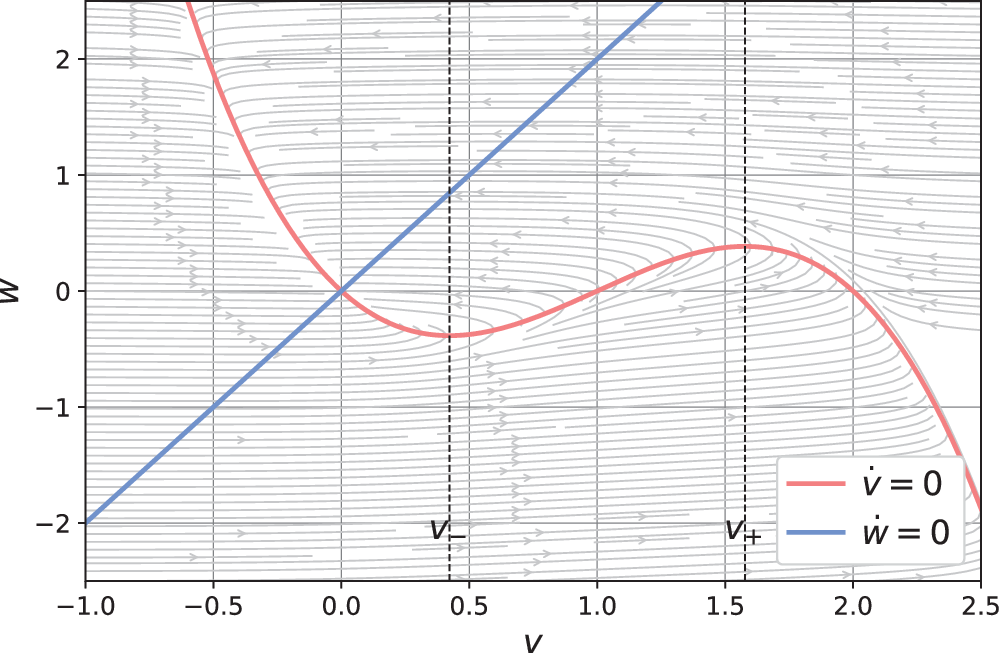}
    \caption{FN phase-plane plot in the excitable regime, with the fast nullcline depicted in red and the slow nullcline in blue. The parameters are $I_0= -1, \alpha = 0.4, \varepsilon = 0.1$, and $\gamma = 1.0$.
The unique fixed point lies on the left branch of the cubic fast nullcline and is stable. The dashed vertical
lines mark the turning points $v_-$ and $v_+$.}
    \label{fig1}
\end{figure}

In the case of the FN model, $f(v)$ is taken to be a cubic function of $v$:
\begin{equation}
\label{FN}
f(v) = v(1-v)(v-\alpha), \qquad 0<\alpha<1.
\end{equation}
 Suppose that the $v$ and $w$ nullclines intersect at a unique point, corresponding to a unique equilibrium of the deterministic system. If the intersection point lies on the left-hand branch of the cubic nullcline, then the system is said to be excitable, since a large perturbation of the stable equilibrium is needed to induce a fast transition to the right-hand branch. The latter then results in a single large excursion in phase space before returning to the rest state, see Fig. \ref{fig1}. Within the neural context, such an excursion represents the firing of an action potential. On the other hand, when the nullcline intersection lies on the middle branch,
the equilibrium is unstable and is surrounded by a stable limit cycle, which represents sustained
periodic spiking, that is, repetitive action potentials. The system is now said to operate in the oscillatory regime.

\begin{figure}[htbp]
    \centering
                      \includegraphics[width=\linewidth]{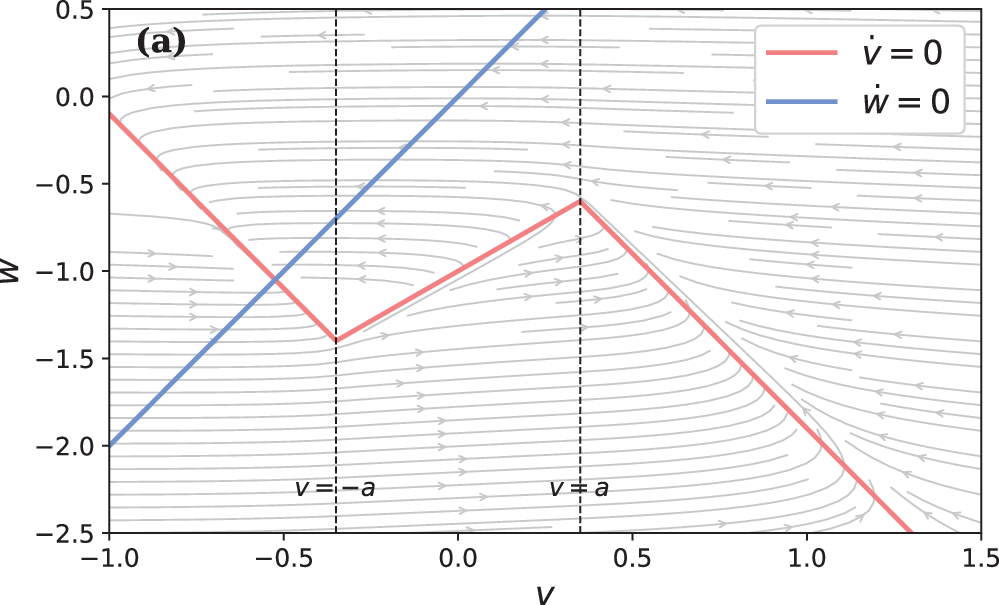}
                       \includegraphics[width=\linewidth]{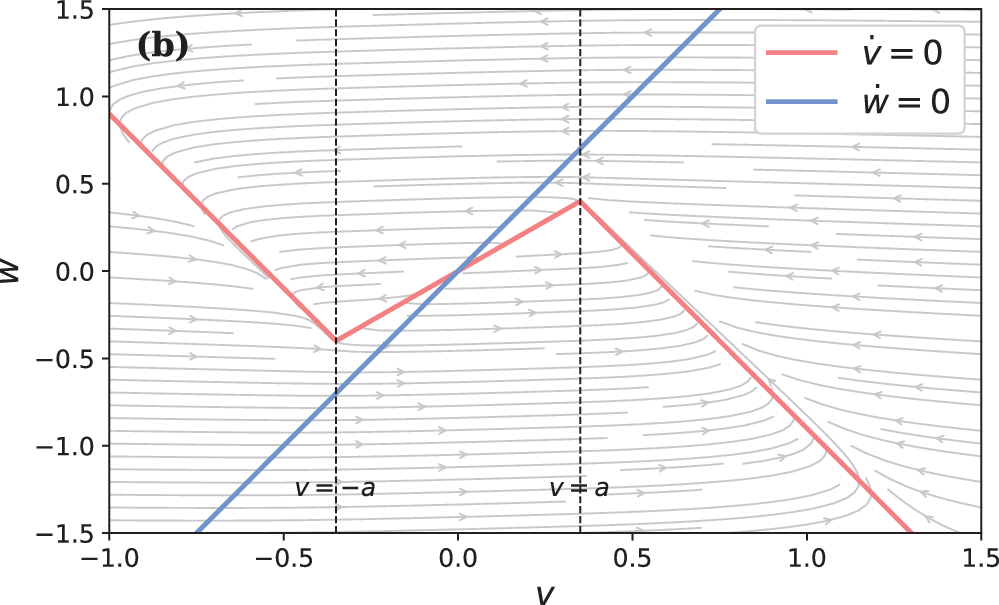}
\caption{McKean phase-plane plots in (a) the excitable regime with $I_0=-1$ and (b) the oscillatory regime with $I_0=0$. The fast nullcline is depicted in red and the slow nullcline in blue. The remaining parameters are $a=0.35$, $\varepsilon =0.1$, $\gamma=0.5$, $m_{\pm }=-2$ and $m_0=1.14$.} 
    \label{fig2}
\end{figure}

\begin{figure}[htbp]
    \centering
        \includegraphics[width=\linewidth]{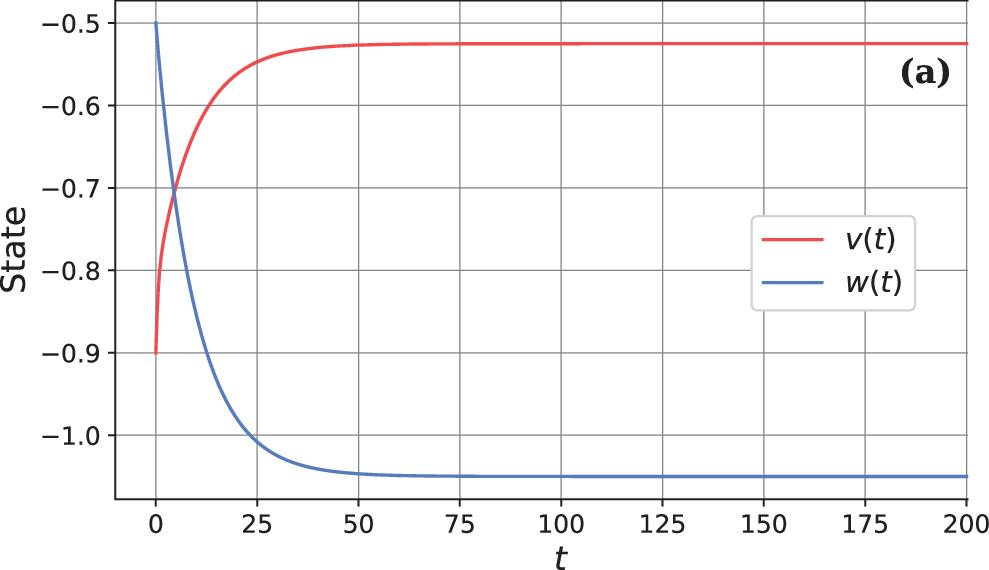}
               \includegraphics[width=\linewidth]{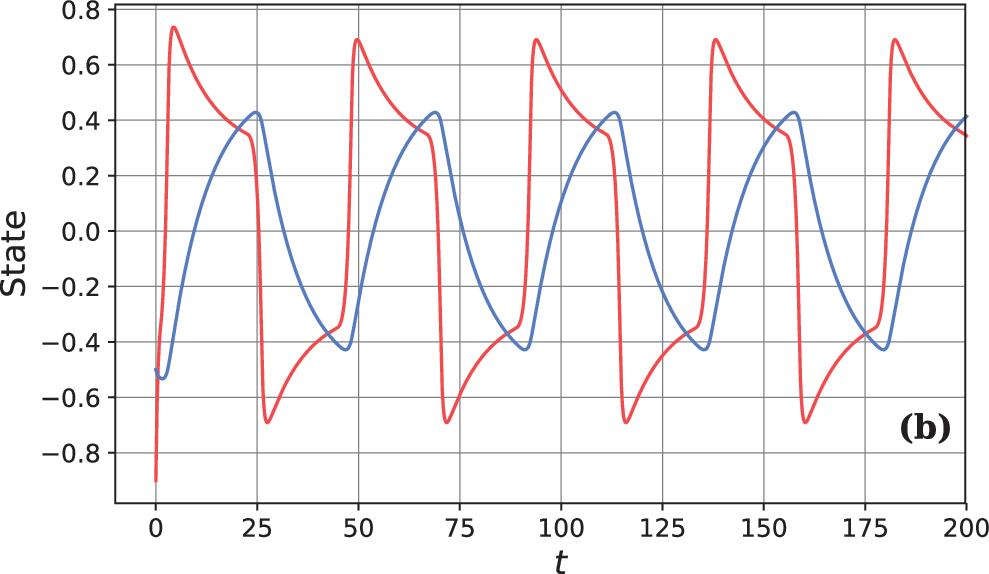}
           \caption{Numerical time series plots of the slow and fast variables in (a) the excitable regime with $I_0=-1$ and (b) the oscillatory regime with $I_0=0$. Other parameter values are the same as Fig. \ref{fig2}.
    \label{fig3}}
\end{figure}

In this paper we replace the cubic nonlinearity by the piecewise linear function
\begin{equation}
\label{MK}
f(v)=
\begin{cases}
m_- v + b_-, &  v\in {\mathcal I}_- :=(-\infty,-a) ,\\
m_0 v+b_0, &v\in {\mathcal I}_0:=[-a,a],\\
m_+ v + b_+, & v\in {\mathcal I}_+:=(a,\infty),
\end{cases}
\end{equation}
with
\begin{equation}
b_0=0,\quad b_\pm := \pm a(m_0-m_\pm).
\end{equation}
We assume $m_-<0, \:m_+<0$ and $0 <m_0$, so that $f$ has two outer branches with negative slope and a middle branch with positive slope, and we choose $b_\pm$ so that $f$ is continuous at the transition points $v=\pm a$. Eqs. (\ref{eqvw}) with $f(v)$ given by (\ref{MK}) yields the McKean model. The latter also exhibits both excitable and oscillatory regimes for appropriate parameter choices. This is illustrated by the phase-plane plots in Fig. \ref{fig2} and the corresponding time series plots in Fig. \ref{fig3}.

A powerful tool for analyzing the qualitative behavior of the FN and McKean models is {\em geometric singular perturbation theory}\cite{Fenichel79,Jones95,Wech20}. First, taking the singular limit $\varepsilon \rightarrow 0$ in Eqs. (\ref{eqvw}) yields $dw(t)/dt=0$ so that on the fast timescale we can treat $w(t)$ as a constant parameter $\overline{w}$. The fast dynamics thus reduces to the {\em layer problem}
\begin{equation}
\label{lay}
\frac{dv}{dt} = f(v) - \overline{w} + I_0.
\end{equation}
 For fixed $\overline{w}$ we can determine the fixed points of the 1D system
and their stability, exploring the onset of bifurcations as the parameter $\overline{w}$ is varied. The dynamics of the slow variable is then obtained by performing the rescaling $\tau = \varepsilon t$ and taking the limit \(\varepsilon \to 0^+\). This yields the {\em reduced problem}
\begin{align}
\label{sf}
0= f(v) - w + I_0,\qquad \frac{dw}{d\tau}=v - \gamma w.
\end{align}
We effectively have a 1D dynamical system for the slow variable $w$ in which the fast variable $v$ adjusts instantaneously to changes in the slow process according to an algebraic equation that constrains the fast process to lie on the nullcline (slow manifold) $f(v)-w+I_0=0$. For deterministic systems, the geometrical aspect of the analysis refers to the fact that it is necessary to take into account the detailed structure of the slow manifold in order to match the reduced and layer solutions. With regards the phase-plane diagrams shown in Figs. \ref{fig1} and \ref{fig2}, whenever the layer solution for $v$ reaches an extremum of the left-hand or right-hand branch of the $v$-nullcline, the slow dynamics has nowhere else to go. The subsequent motion involves
the rapid evolution of $v$ on horizontal lines that terminate at points of intersection on the opposite branch.

\subsection{Stochastic model}

We now formulate a stochastic version of the slow-fast system (\ref{eqvw}) by adding both Gaussian white noise and Poisson resetting to the dynamics of the fast variable. In particular, we assume that the fast variable instantaneously resets to a fixed value $v_r$ at the sequence of times $\{T_{n},n\geq 1\}$ generated by a homogeneous  Poisson process $N(t)$ with rate $r$. That is, the inter-reset intervals $\tau_{n}=T_{n+1}-T_{n}$ are exponentially distributed random variables with 
\begin{equation}
{\mathbb P}[\tau_{n}\in [\tau,\tau+d\tau]]=r\e^{-r\tau}d\tau.
\end{equation}
Moreover, $N(t)$ is right-continuous which means that $N(T_n^-)=n-1$ whereas $N(T_n)=n$. (Throughout the paper we use the notation $t^-$ to mean approaching $t$ from the left-hand side, that is, $t^-=\lim_{\delta \rightarrow 0}(t-\delta)$ where $\delta $ is a positive constant.)
The resulting SDE takes the form
\begin{subequations}
\label{SDE}
\begin{align}
dV(t) &= [f(V(t)) - W(t) + I_0]\,dt + \sqrt{2D}\,dB(t) \nonumber \\
&\quad + [v_r - V(t^-)]\,dN(t)  , \\
dW(t) &= \varepsilon (V(t) - \gamma W(t)) dt,
\end{align}
\end{subequations}
where $B(t)$ is a standard Brownian motion with $\langle B(t)\rangle =0$ and
\begin{equation}
\langle B(t)B(s)\rangle =\min \{t,s\},
\end{equation}
$D$ is an effective non-dimensionalized diffusivity that specifies the strength of the Gaussian noise, and
\begin{equation}
\label{dNt}
dN(t):=h(t)dt=\sum_{n\geq 1}\delta(t-{T}_n)dt.
\end{equation}
The second line of Eq. (\ref{SDE}a) implements the jumps from $V(t^-)$ to $v_r$ at the resetting times $t=T_n$.

Note the inverse resetting rate $r^{-1}$ introduces a new time-scale into the problem and the effects of resetting will depend on how $r^{-1}$ compares to the slow time-scale $\varepsilon$. (Recall that we are working in dimensionless units.) For the moment we will take $r=O(1)$ so that resetting operates on the fast timescale. We will consider the regime in which resetting occurs on the slow timescale in Sect. VI.

We can exploit the separation of timescales to develop stochastic analogs of the layer problem and reduced problem \cite{Bressloff25}. First, in the limit $\varepsilon \to 0$ we have $dW(t)=0$ so that we can set $W(t) = \overline w \in \mathbb{R}$. This yields a 1D layer problem with resetting and diffusion:
\begin{align}
dV(t) &= [f(V(t)) - \overline w + I_0]\,dt + \sqrt{2D}\,dB(t) \nonumber \\
&\quad + [v_r - V(t^-)]\,dN(t).
\end{align}
Introduce the probability density
\begin{equation}
p(v,t)={\mathbb E}[\delta(V(t)-v)],
\end{equation}
where expectation is taken with respect to both the Gaussian white noise and Poisson resetting. Using a generalized It\^o's lemma, it can be shown that $p(v,t)$ satisfies the Fokker-Planck equation (see Appendix B of Ref. \onlinecite{Bressloff25})
\begin{align}
\frac{\partial p(v,t)}{\partial t} &= -\frac{\partial}{\partial v} \big[(f(v) - \overline w + I_0)p(v,t)\big] + D \frac{\partial^2 p(v,t)}{\partial v^2} \nonumber \\
&\quad - r p(v,t) + r \delta(v - v_r) \label{eq:Fokker_Planck},
\end{align}
with initial condition $p(v, 0) = \delta(v - v_0)$, corresponding to the process being initialised at the point $v_0$. The latter is often identified with the reset point $v_r$. Physically speaking, the first and second terms on the right-hand side of Eq. (\ref{eq:Fokker_Planck}) involve the deterministic and Fickian fluxes, whereas the final two terms represent the loss and gain of probability at the reset point $v_r$ due to resetting.

Suppose that there exists a unique NESS $p^*$ such that 
\begin{equation}
\lim_{t\rightarrow \infty}p(v,t)= p^*(v|\overline{w}).
\end{equation}
It is convenient to make the dependence of the NESS on $\overline{w}$ explicit. Since the probability density converges on the fast timescale, we can exploit this in determining the dynamics of $W(t)$ on the slow timescale $\tau = \varepsilon t$. However, in contrast to the reduced problem of the deterministic system, we can no longer assume that the fast variable is constrained to lie on the slow manifold. Following Ref. \onlinecite{Bressloff25}, we replace the fast variable by its average with respect to the NESS. Thus, the reduced problem becomes
\begin{equation}
\label{slow}
    \frac{dw}{d\tau} = \mathbb{E}[V| w(\tau)] - \gamma w(\tau),
\end{equation}
where
\begin{equation}
\label{sfav}
\mathbb{E}[V| w]
=
\int_{\mathbb{R}} v\, p^*(v |w)\, dv.
\end{equation}
Note that the solution of Eq. (\ref{slow}) determines the corresponding slow variation of the NESS according to $p^*(v,\tau)=p^*(v|w(\tau))$.
It follows that the slow variable converges to a stable fixed point $w^*$ given by a root of the equation
\begin{equation}
\label{FPw}
\mathbb{E}[V |w^*] = \gamma w^*,
\end{equation}
such that $d\mathbb{E}[V |w] /dw<0$ at $w=w^*$, and the full 2D system reaches the stationary distribution \(p^*(v |w^*)\). This result holds irrespective of whether the underlying deterministic version of the model operates in an excitable or oscillatory regime, and which branch of the slow manifold $v_r$ is located.

In Ref. \onlinecite{Bressloff25} we analyzed the stochastic layer and reduced problems for the FN model with resetting but without diffusion ($D=0$). In particular, we were able to obtain an explicit expression for the NESS. However, the analysis becomes considerably more difficult when $D>0$. Therefore, we will proceed by working with the McKean model.

\setcounter{equation}{0}
\section{Analysis of the non-diffusive case ($D=0$)}

We first analyze the stochastic McKean model given by Eqs. (\ref{SDE}), with $f(v)$ defined by Eq. (\ref{MK}), in the non-diffusive case $D=0$. We will subsequently use our analytical results as a reference point for the diffusive case in Sects. IV and V.
For the sake of illustration, we assume that the underlying deterministic system without resetting operates in the excitable regime as shown in Figs. \ref{fig2}(a) and \ref{fig3}(a). (Fast resetting eliminates the limit cycle in the oscillatory regime, resulting in similar behavior to the excitable case. On the other hand, noisy oscillations are observed when $r=O(\varepsilon)$, see Sect. VI.)

\subsection{NESS for the layer problem}
 The stationary FP equation (\ref{eq:Fokker_Planck}) with $D=0$ reduces to the first-order equation
\begin{equation}
0 =
-\frac{\partial}{\partial v}\Big[g(v)\,p^*(v)\Big]
-r\,p^*(v )
+r\,\delta(v-v_r),
\label{eq:stat_D0}
\end{equation}
where we have defined
\begin{equation}
\label{gv}
g(v):=f(v)-\overline w + I_0.
\end{equation}
Introducing the stationary probability flux
\begin{equation}
J(v):=g(v)\,p^*(v),
\label{eq:def_flux}
\end{equation}
we can rewrite Eq. \eqref{eq:stat_D0} as
\[
J'(v)=-r\,p^*(v)=-r\,\frac{J(v)}{g(v)}
\label{eq:J_ode}
\]
for all $v\neq v_r$, which can be rearranged as
\begin{equation}
\frac{J'(v)}{J(v)}=-\frac{r}{g(v)}.
\end{equation}
For the McKean model, $ g(v)$ is piecewise linear and continuous, with
\begin{equation}
    g(v)=
\begin{cases}
 g_-(v). & v\in {\mathcal I}_-  ,\\ 
 g_0(v) , & v\in {\mathcal I}_0 ,\\  
g_+(v), & v\in {\mathcal I}_+  ,
\end{cases}
\label{gvpw}
\end{equation}
and
\begin{equation}
 g_i(v) =m_i v + b_i+I_0-\overline w.
 \end{equation}
 Hence, we have the piecewise general solution
 \begin{equation}
 J(v)= C_i\Gamma_{i}(v),\quad v \in {\mathcal I}_i,
 \end{equation}
 where
 \begin{align}
    \Gamma_i(v)&=\exp\!\left(-r\int^v \frac{du}{g_{i}(u)}\right)\nonumber  \\
    &=\,|m_{i} v + b_{i}+I_0-\overline w|^{-r/m_{i}}.
    \end{align}
   The corresponding general solution for the NESS is,
\begin{align}
\label{solpstar}
    p^*(v)&=\frac{C_{i} \Gamma_{i}(v)}{g_{\i}(v)} ,\quad v\in {\mathcal I}_{i}.
    \end{align}

The above construction has to be modified when taking into account the singularity at $v=v_r$, which leads to a jump discontinuity in the flux. Suppose, for the sake of illustration, that $-a<v_r<a$ so the reset point lies on the middle branch. (The construction of the solution when $|v_r|>a$ can be developed along similar lines.) We now write
\begin{subequations}
\label{solJ}
\begin{align}
    J(v)&=C_{\pm} \Gamma_{\pm}(v), \quad v\in {\mathcal I}_{\pm},\\
     J(v)&=C_{<} \Gamma_{0}(v), \quad -a<v<v_r,\\
      J(v)&=C_{>} \Gamma_{0}(v), \quad v_r<v<a,
    \end{align}
    \end{subequations}
    and
    \begin{subequations}
\label{solJp}
\begin{align}
    p^*(v)&=\frac{C_{\pm} \Gamma_{\pm}(v)}{g_{\pm}(v)}, \quad v\in {\mathcal I}_{\pm},\\
     p^*(v)&=\frac{C_{<} \Gamma_{0}(v)}{g_0(v)}, \quad -a<v<v_r,\\
      p^*(v)&=\frac{C_{>} \Gamma_{0}(v)}{g_0(v)}, \quad v_r<v<a.
    \end{align}
    \end{subequations}
  Integrating Eq. \eqref{eq:stat_D0}
across a small region containing the reset point and taking the limit \(\delta\to0^+\), we obtain
\begin{align}
&\lim_{\delta\to0^+}\int_{v_r-\delta}^{v_r+\delta}\big(-J'(v)\big)\,dv
 \\
&=
\lim_{\delta\to0^+}\int_{v_r-\delta}^{v_r+\delta} r\,p^*(v)\,dv- \lim_{\delta\to0^+}\int_{v_r-\delta}^{v_r+\delta} r\,\delta(v-v_r)\,dv,
\nonumber 
\end{align}
which yields the flux condition
\begin{equation}
  J(v_r^+)-J(v_r^-)\equiv  (C_>-C_<)\Gamma_0(v_r)=r  . \label{eq:flux_cond}
\end{equation}
On the other hand, the flux is continuous at the piecewise transition points $v=\pm a$, which implies that
\begin{equation}
C_-\Gamma_-(-a)=C_<\Gamma_0(-a),\quad C_+\Gamma_+(a)=C_>\Gamma_0(a).
\label{eq:flux_kinks}
\end{equation}

 \begin{figure}[t!]
\centering
\includegraphics[width=\linewidth]{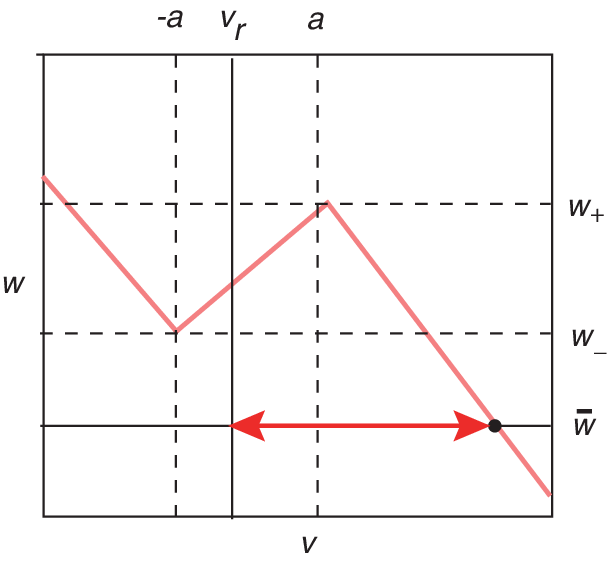}
\caption{Schematic diagram illustrating the support of the layer dynamics when $-a<v_r<a$. The intercept of the $v$-nullcline $w=f(v)+I_0$ with the horizontal line $w=\overline w$ is unique when $\overline w < w_-$, where $w_{\pm}$ are the turning points of the $v$-nullcline. In this scenario the support of $v$ is the interval linking the reset point $v_r$ to $v_+$.}
\label{fig4}
\end{figure}

We currently have three conditions for the four unknown coefficients $C_{\pm},C_<$ and $C_>$. One cannot obtain a fourth condition by imposing the unit normalization $\int_{\R}p^*(v)dv=1$ since the latter is automatically satisfied. However, we haven't yet taken into account the fact that the support of the NESS with respect to $v$ is finite (in the absence of diffusion). The latter is determined by the location of the reset state $v_r$ relative to the roots of \(g(v)\), which are the points of intercept between the $v$-nullcline $w=f(v)+I_0$ and the horizontal line $w=\overline w$. First suppose that \(g(v)\) has a single root $v_+$ that lies on the right-hand branch and hence $v_r<v_+$ (since we have assumed that $v_r$ lies on the middle branch), see Fig. \ref{fig4}. As $v_+$ is a stable fixed point of Eq. (\ref{lay}), it follows that after each reset the dynamics will transition from $v_r$ back to $v_+$. Hence, the support of $v$ will be $[v_r,v_+) $ . Returning to the flux condition (\ref{eq:flux_cond}), we note that no trajectory can cross $v_r$ from the right, which means that $J(v_r^-)=0$. Hence, $C_<=0$ and 
\begin{equation}
C_>=\frac{r}{\Gamma_0(v_r)}.
\end{equation}
The continuity conditions (\ref{eq:flux_kinks}) then show that $C_-=0$ and
\begin{equation}
C_+=\frac{C_>\Gamma_0(a)}{\Gamma_+(a)}=\frac{r\Gamma_0(a)}{\Gamma_+(a)\Gamma_0(v_r)}.
\end{equation}
Also note that $g(v)>0$ for all $v\in [v_r,v_+)$, which ensures that $p(v)>0$.

\begin{figure}[t]
    \centering
            \includegraphics[width=\linewidth]{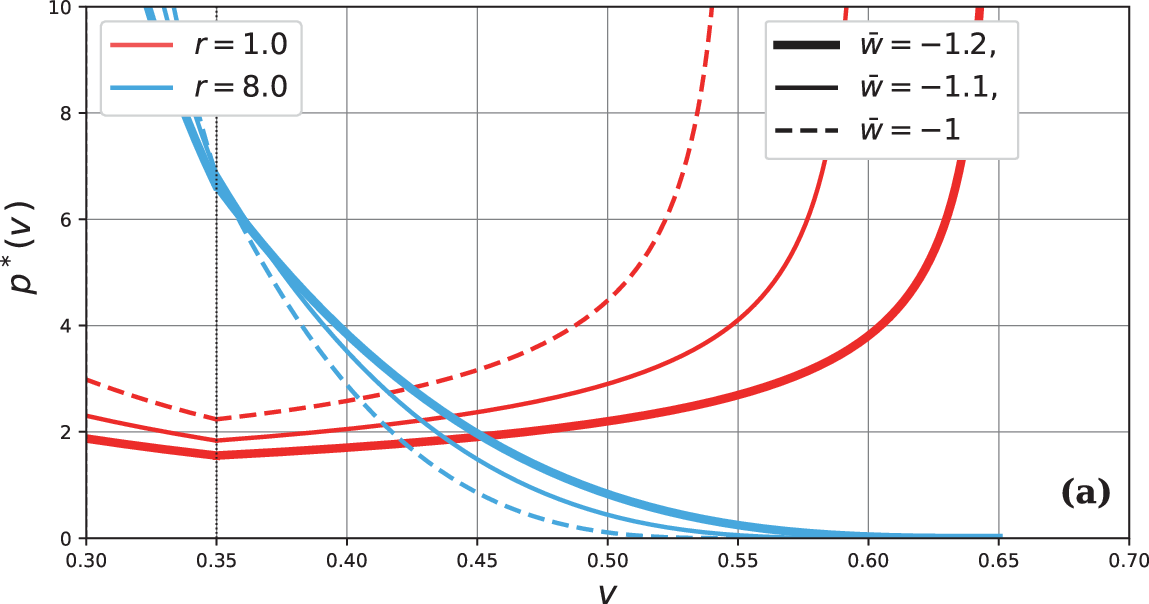}
            \includegraphics[width=\linewidth]{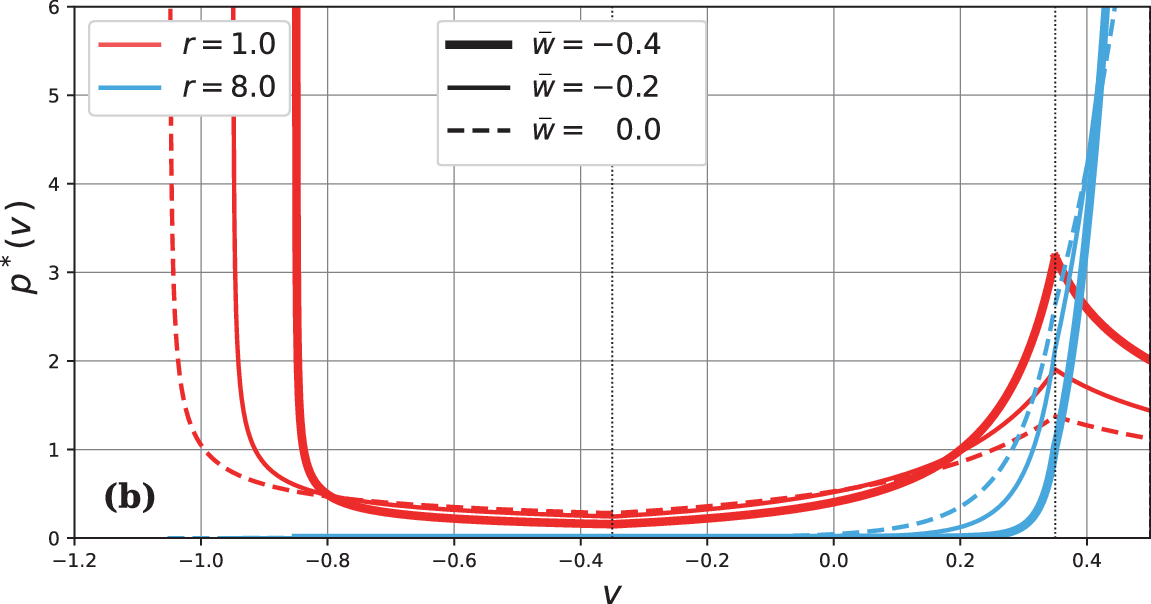}
          \includegraphics[width=\linewidth]{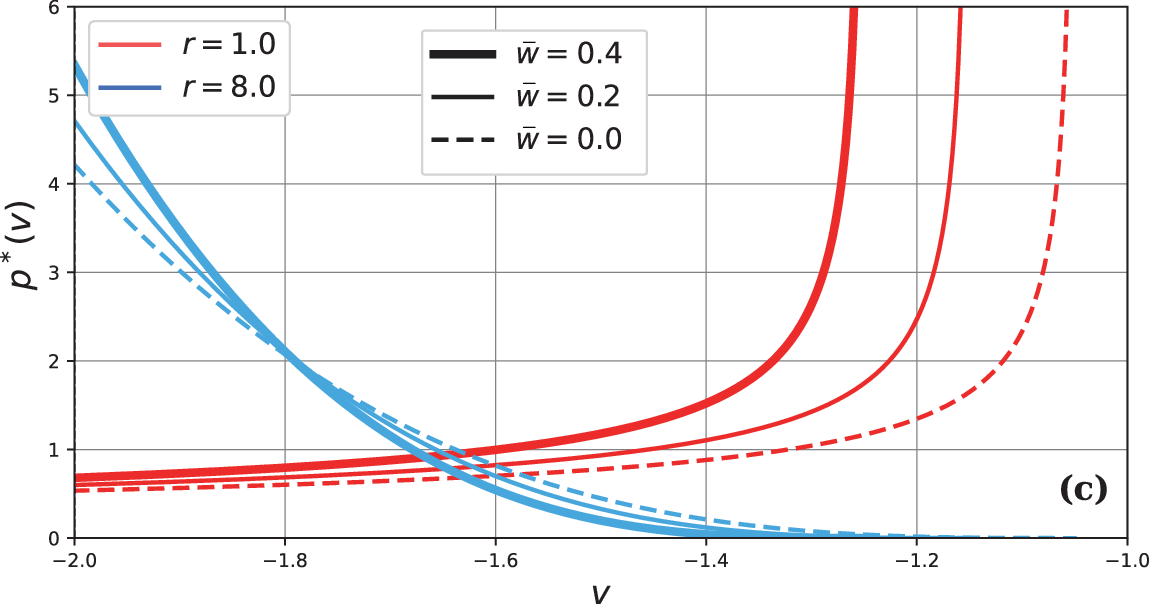}
    \caption{Plots of \(p^{*}(v)\) with resetting in the non-diffusive excitable regime for different values of the frozen slow variable $\overline w$ and the resetting rate $r$. Other parameters are $I_0=-1$, $a=0.35$, $m_{\pm }=-2$ and $m_0=1.14$. (a) $v_r=0.3$ (resetting to the middle branch), (b) $v_r=0.5$ (resetting to the right-hand branch), and (c) $v_r=-2$ (resetting to the left-hand branch). As the resetting rate increases, probability accumulates more strongly near the reset point \(v_r\) rather than near the fixed point $v_+$.}
    \label{fig5}
\end{figure}

A similar analysis can be performed for $\overline w>w_+$, where the only root $v_-$ of $g(v)$ lies on the left-hand branch. In this case, $v\in (v_-,v_r]$, $C_>=C_+=0$, 
\begin{equation}
C_<=-\frac{r}{\Gamma_0(v_r)},
\end{equation}
 and
\begin{equation}
C_-=\frac{C_<\Gamma_0(-a)}{\Gamma_-(-a)}=-\frac{r\Gamma_0(-a)}{\Gamma_-(-a)\Gamma_0(v_r)}.
\end{equation}
Here we have $g(v)<0$ for all $v\in (v_-,v_r]$, which ensures that $p(v)>0$ even though $C_<,C_- <0$. Finally, suppose that $w_-<\overline w < w_+$. Now the function $g(v)$ has three roots $v_{\pm},v_0$ with $v_{\pm}$ stable equilibria of the ODE (\ref{lay}) and $v_0$ an unstable equilibrium. There are then two possibilities generically: (i) $v_0<v_r$ so that $v_r$ lies in the basin of attraction of $v_+$ and $v\in [v_r,v_+)$; (ii) $v_0> v_r$ so that $v_r$ lies in the basin of attraction of $v_-$ and $v\in (v_-,v_r]$. These two cases can then be analyzed as before. 

Example plots of \(p^*(v)\) for different values of \( \overline w\) and the resetting rate $r$ are shown in Figs. \ref{fig5} for $v_r$ located on (a) the middle branch, (b) the right-hand branch and (c) the left-hand branch. Observe the change in the support of $p^*(v)$ as \( \overline w \) changes. Moreover, the point of accumulation of the NESS can change as the resetting rate varies.

\subsection{Averaged slow dynamics}
\label{sec:Averaged_Slow_Dynamics}

\begin{figure}[b!]
    \centering
               \includegraphics[width=\linewidth]{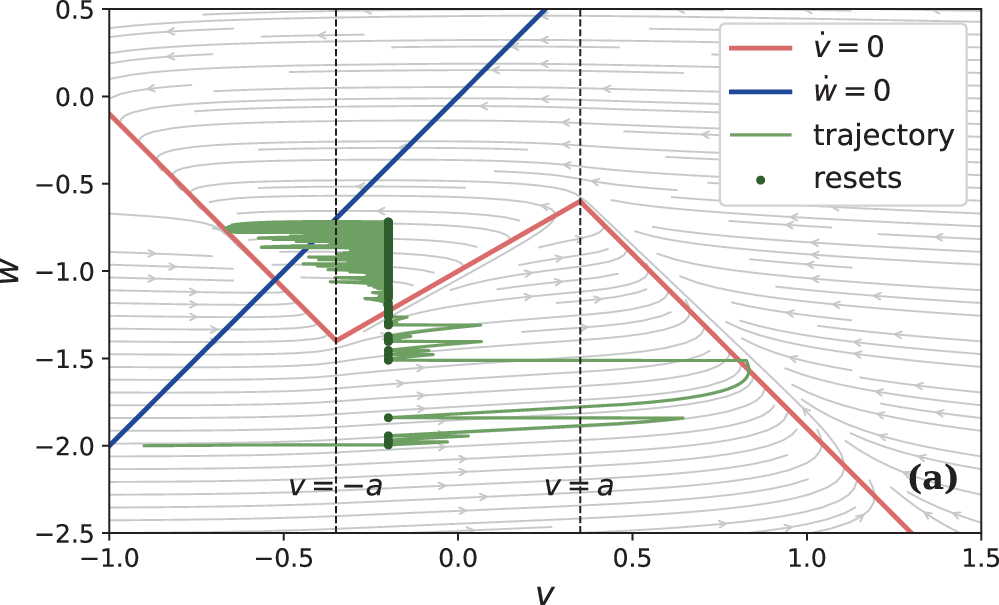}
               \includegraphics[width=\linewidth]{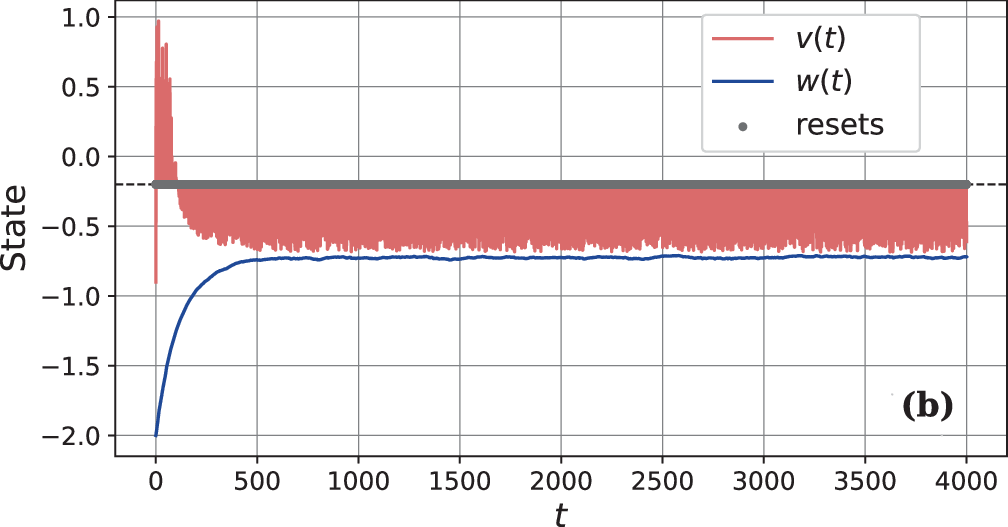}
           \caption{Numerical plots illustrating convergence to a stable equilibrium of the slow dynamics with resetting in the non-diffusive case. We take  \(r=3\), $v_r = -0.2$ and the initial conditions $v(0)=-0.9,w(0)=-2$. (a) Trajectory in the phase-plane for $\varepsilon =0.1$ showing resetting of the fast variable combined with a slowly increasing slow variable. Note the switch in the support of the fast variable as it crosses the middle branch. (b) Corresponding time series plots of the fast variable $v(t)$ (red) and the slow variable $w(t)$ (blue) with $\varepsilon=0.01$. Other parameters are $I_0=-1$, $a=0.35$, $\gamma=0.5$, $m_{\pm }=-2$ and $m_0=1.14$.}
    \label{fig6}
\end{figure}

\begin{figure}[t!]
    \centering
    \includegraphics[width=\linewidth]{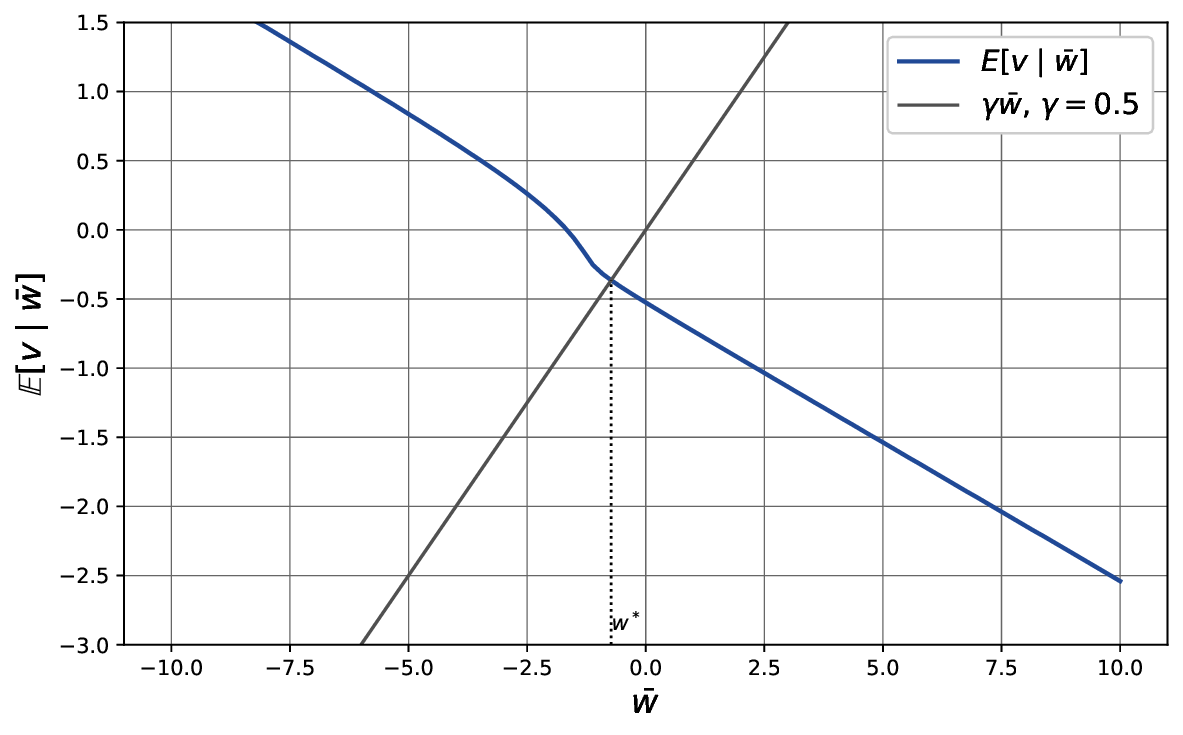}
    \caption{Numerically evaluated plot of \( \mathbb{E}[v|w] \) as a function of $w$. Same parameter values as Fig. \ref{fig6}. The intersection with the straight line $y=\gamma w$ yields the fixed point \(w^* \approx -0.7286\).}
    \label{fig7}
\end{figure}

So far we have calculated the NESS $p^*(v|\overline w)$ for a fixed value of the slow variable $\overline w$ and shown that the NESS is restricted to a finite interval. (It is now convenient to make the dependence of the NESS on $\overline w$ explicit.) The corresponding slow dynamics is then given by Eqs. (\ref{slow}) and (\ref{sfav}), after taking $\overline w\rightarrow w(\tau)$. We thus expect the slow variable to converge to a stable fixed point $w^*$ satisfying Eq. (\ref{FPw}). For the sake of illustration, 
suppose that the system starts in a state $(v(0),w(0))$ with $w(0)<w_-$ and resetting to the middle branch along the lines of Fig. \ref{fig4}. We also assume that the underlying deterministic system without resetting operates in the excitable regime with a resting equilibrium state on the left-hand branch of the $v$-nullcline, see Fig. \ref{fig2}(a).  In Fig. \ref{fig6} we show numerical simulations of the full 2D system. This shows fast resetting to the middle branch, a slowly increasing recovery variable $w(t)$ that induces a switching of support as $(v_r,w)$ crosses the middle branch, and convergence of $w(t)$ to a fixed point $w^*$ in the large time limit. The time-dependent approach to the fixed point can be understood as follows. Let
\begin{equation}
v_{\rm av}(\tau) ={\mathbb E}[V|w(\tau)]=\int_{\Sigma(\tau)} p(v|w(\tau))vdv,
\end{equation}
where $\Sigma(\tau)$ is the support of the NESS when $w=w(\tau)$. Also set
\begin{equation}
g(v,\tau)=f(v)+I_0-w(\tau),
\end{equation}
and denote the roots of $g(v,\tau)$ by $v_{\pm}(\tau),v_0(\tau)$ (when they exist). Initially the fast variable increases towards a point $v_+(0)$ on the right-hand branch after each reset. The support of the NESS is $\Sigma(0)=[v_r,v_+(0))$ and lies below the $w$-nullcline. It follows that $v_{\rm av}(\tau) > \gamma w$ and thus $dw/d\tau >0$. Since $w(\tau)$ is an increasing function of the slow time $\tau$, we have the following sequence of events as seen in Fig. \ref{fig6}. First, $w(\tau)$ crosses $w_-$ from below and the corresponding function $g(v,\tau)$ develops three roots $v_{\pm}(\tau),v_0(\tau)$ with $v_0(\tau) <v_r$. Second, the middle root crosses $v_r$ from the left and the support of the NESS switches from $[v_r,v_+(\tau))$ to $(v_-(\tau),v_r]$. Third, since $v_-(\tau)$ is a monotonically decreasing function of $\tau$, the support of the NESS eventually includes a region above the $w$-nullcline, resulting in a decreasing average $v_{\rm av}(\tau)$. Fourth, $\lim_{\tau\rightarrow \infty} v_{\rm av}(\tau) =\gamma w^*$ and the slow variable reaches an equilibrium $w^*$.

\begin{figure}[t!]
    \centering
        \includegraphics[width=\linewidth]{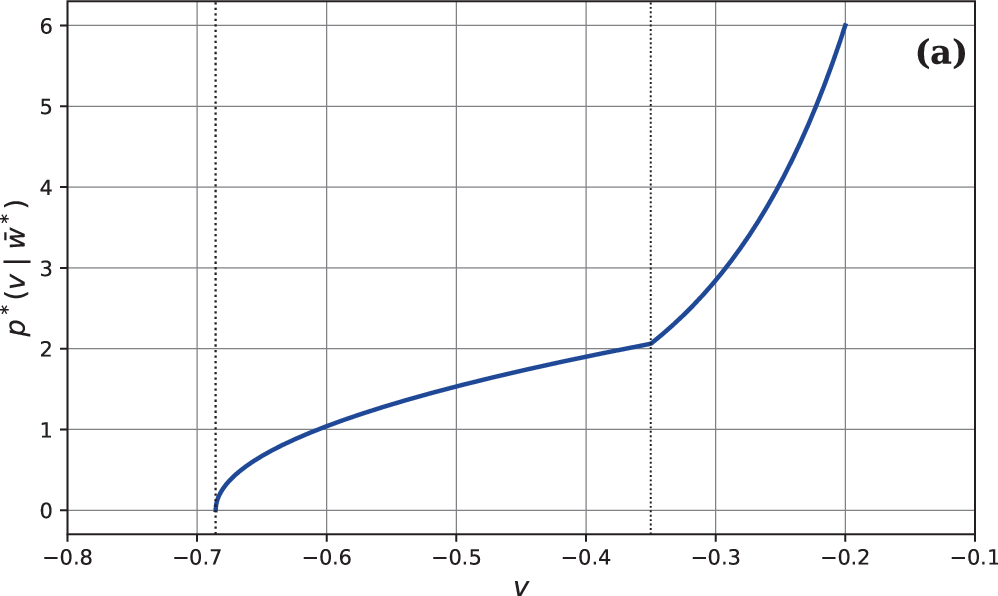}
               \includegraphics[width=\linewidth]{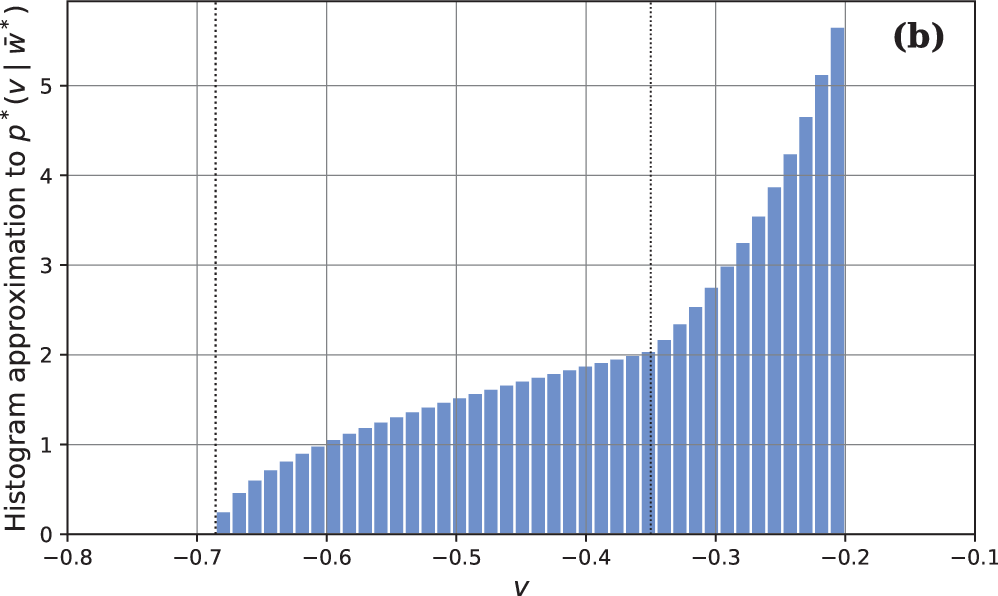}
           \caption{The NESS $p^*(v|w^*)$ at the equilibrium $w^*$ of the slow dynamics. Same parameter values as Fig. \ref{fig6} and \ref{fig7}. (a) Plot of the analytically determined stationary solution at the slow fixed point \(w^* \approx -0.7286\). (b) A histogram depicting the numerically depicted probability density.}
    \label{fig8}
\end{figure}

In order to check that $w^*$ satisfies Eq. (\ref{FPw}), we numerically evaluate ${\mathbb E}[V|w]$ as a function of $w$ using the analytical expression for $p^*(v|w)$ obtained in Sect. III.A. The resulting function of $w$ is plotted in Fig. \ref{fig7} for the same parameters as Fig. \ref{fig6}. It can be seen that for this particular example,  ${\mathbb E}[V|w]$ is a strictly monotonically decreasing function of $w$ that has a unique intercept with the straight line $y=\gamma w$, which corresponds to a stable equilibrium $w^*$. Moreover, we find that $w^*\approx -0.7286$, which agrees very well with the asymptote shown in Fig. \ref{fig6}(b). Using this fixed point, the resulting NESS of the full 2D system is $p^*(v,w)=p^*(v|w^*)\delta(w-w^*)$. The stationary density $p^*(v|w^*)$ corresponding to the same setup as Figs. \ref{fig6} and \ref{fig7} is given in Fig. \ref{fig8}. The resulting probability distribution clearly shows that probability accumulates near the reset point.

\setcounter{equation}{0}
\section{Analysis of the diffusive case ($D>0$)}

\begin{figure}[t!]
    \centering
        \includegraphics[width=\linewidth]{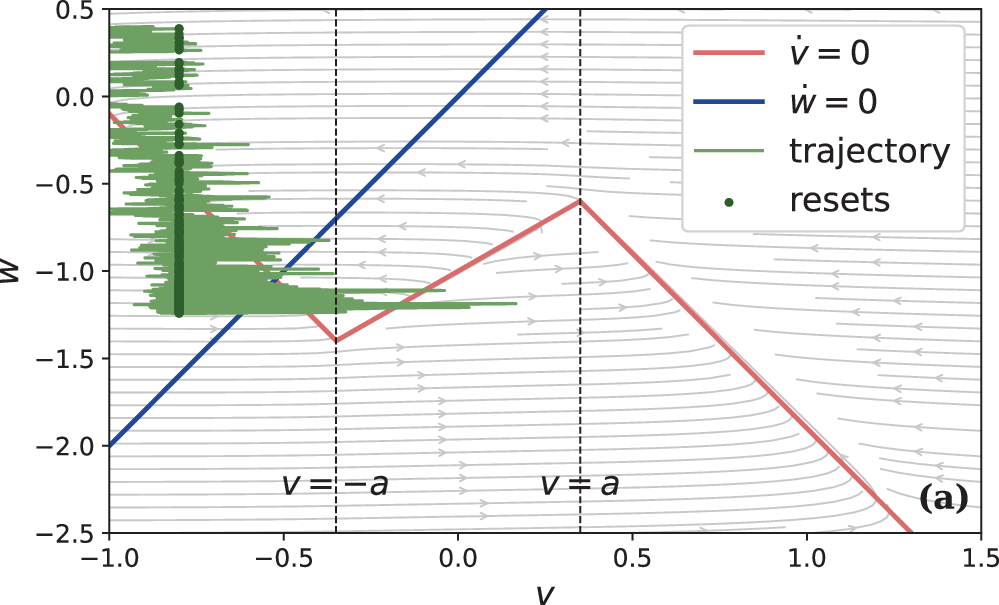}
               \includegraphics[width=\linewidth]{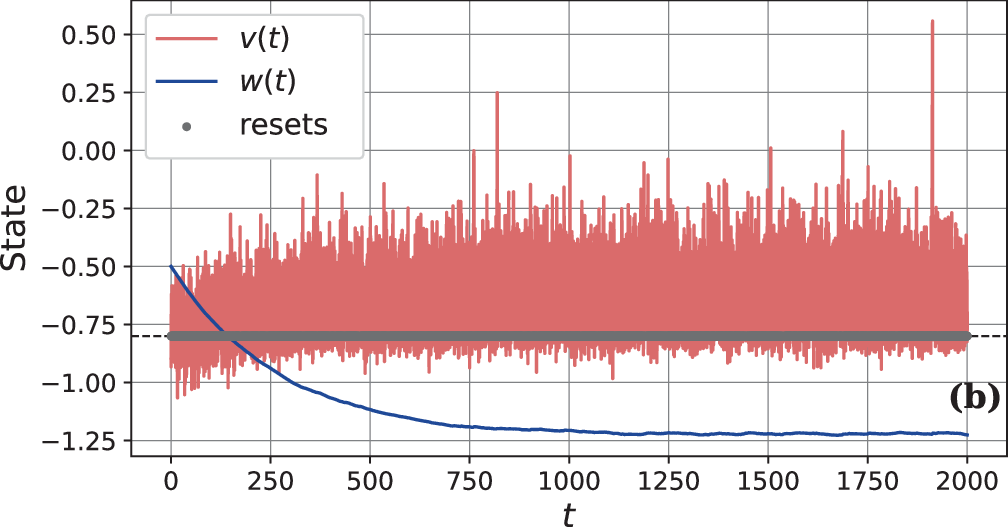}
                      \includegraphics[width=\linewidth]{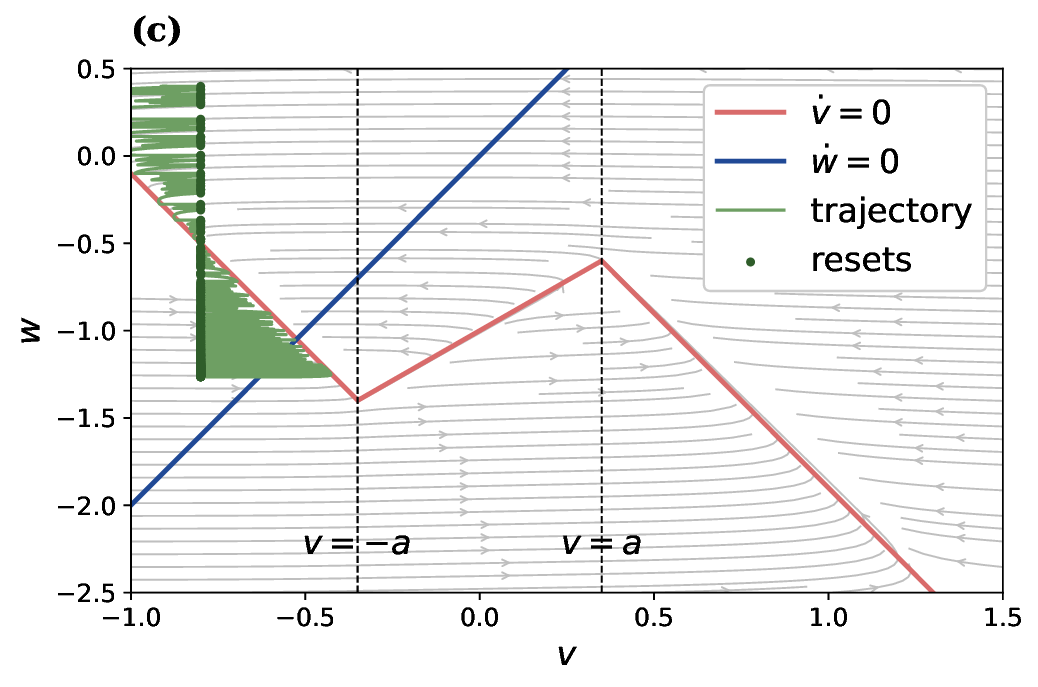}
               \caption{Numerical plots illustrating convergence to a stable equilibrium of the slow dynamics in the presence of both resetting and diffusion. We take $D=0.01$, \(r=2\), $v_r = -0.8$ and the initial conditions $v(0)=-0.9,w(0)=0.4$. (a) Trajectory in the phase-plane for $\varepsilon =0.04$ showing resetting of the fast variable combined with a slowly decreasing slow variable. (b) Corresponding time series plots of the fast variable $v(t)$ (red) and the slow variable $w(t)$ (blue) with $\varepsilon=0.04$. (c) Corresponding phase-plane plot when $D=0$. Other parameters are as in Fig. \ref{fig6}: $I_0=-1$, $a=0.35$, $\gamma=0.5$, $m_{\pm }=-2$ and $m_0=1.14$. }
           \label{fig9}
\end{figure} 

We now extend the slow-fast analysis of the stochastic McKean model to include the effects of diffusion ($D>0$). A numerical simulation of the full system is shown in Fig. \ref{fig9}(a,b) with resetting to the left-hand branch and $w(0)>w_+$. As expected, the slow variable converges to a fixed point $w^*$. However, there are two major differences from the non-diffusive case, which is shown for comparison in Fig. \ref{fig9}(c). First, fluctuations in the fast variable are larger. Second, the support of the fast variable is no longer restricted to a fixed finite interval linking $v_r$ to a stable fixed point of the deterministic system (\ref{lay}). 

\subsection{NESS of the layer problem}
The NESS for $D>0$ satisfies the time-independent version of the second-order FP equation \eqref{eq:Fokker_Planck}:
\begin{align}
\label{NESSD}
0=D\,\frac{\partial^2 p^*(v)}{\partial v^2}-\frac{\partial}{\partial v}\bigg [g(v)p^*(v)\bigg ] 
-r\,p^*(v)
+r\delta(v-v_r)
\end{align}
with $g(v)$ given by Eq. (\ref{gvpw}). Introduce the probability flux
\begin{equation}
J(v)=-D\frac{\partial p^*(v)}{\partial v}+g(v)p^*(v).
\end{equation}
At the transition points $v=\pm a$ we have the density and flux continuity conditions
\begin{equation}
\label{con1}
p^*(a^+)=p^*(a^-),\quad J(a^+)=J(a^-)
\end{equation}
and
\begin{equation}
\label{con2}
p^*(-a^+)=p^*(-a^-),\quad J(-a^+)=J(-a^-).
\end{equation}
On the other hand, integrating Eq. (\ref{NESSD}) across the reset point shows that the density is continuous but there is a jump discontinuity in the flux:
\begin{equation}
\label{con3}
p^*(v_r^+)=p^*(v_r^-),\quad J(v_r^+)-J(v_r^-)=r.
\end{equation}

Suppose that we make the substitution \(p^*(v) = e^{S(v)}u(v)\) in Eq. (\ref{NESSD}) such that
\begin{align*}
\frac{\partial p^*(v)}{\partial v}
&=
S'(v)e^{S(v)}u(v) + e^{S(v)}u'(v), \\
\frac{\partial^2 p^*(v)}{\partial v^2}
&=
S''(v)e^{S(v)}u(v)+[S'(v)]^2 e^{S(v)}u(v)\nonumber \\
&\qquad +2S'(v)e^{S(v)}u'(v)
+\e^{S(v)}u''(v).
\end{align*}
Upon substituting for $p^*(v)$, dividing through by \(e^{S(v)}\) and collecting
derivatives, the stationary FP equation becomes
\begin{align}
\label{u1}
0&=D\,u''(v)
+
\bigl[2DS'(v) - g(v)\bigr]u'(v)\\
&+
\Bigl[
DS''(v) + D[S'(v)]^2 - g(v)S'(v) - (g'(v)+r)
\Bigr]u(v) \nonumber 
\end{align}
for $v\neq v_r$.
We choose \(S(v)\) such that the coefficient of the \(u'(v)\) term
vanishes. That is, we set $2DS'(v) =g(v)$. Eq. (\ref{u1}) then reduces to the form
\begin{align}
\label{u2}
D\,u''(v)
-
\left[
\frac{g'(v)}{2} + r + \frac{g(v)^2}{4D}
\right]u(v)
=0.
\end{align}
Eq. (\ref{gvpw}) implies that $g'(v)=m_i$ for $v\in {\mathcal I}_i$ and
\begin{equation}
\label{SK}
S(v)=
\begin{cases}
S_-(v), & v<-a,\\
S_0(v), & v\in[-a,a],\\
S_+(v), & v>a,
\end{cases}
\end{equation}
with
\begin{align}
S_i(v)=\frac{m_i v^2}{4D} + \frac{(b_i+I_0-\overline w)v}{2D}.
\end{align}

\begin{figure}[b!]
    \centering
        \includegraphics[width=0.9\linewidth]{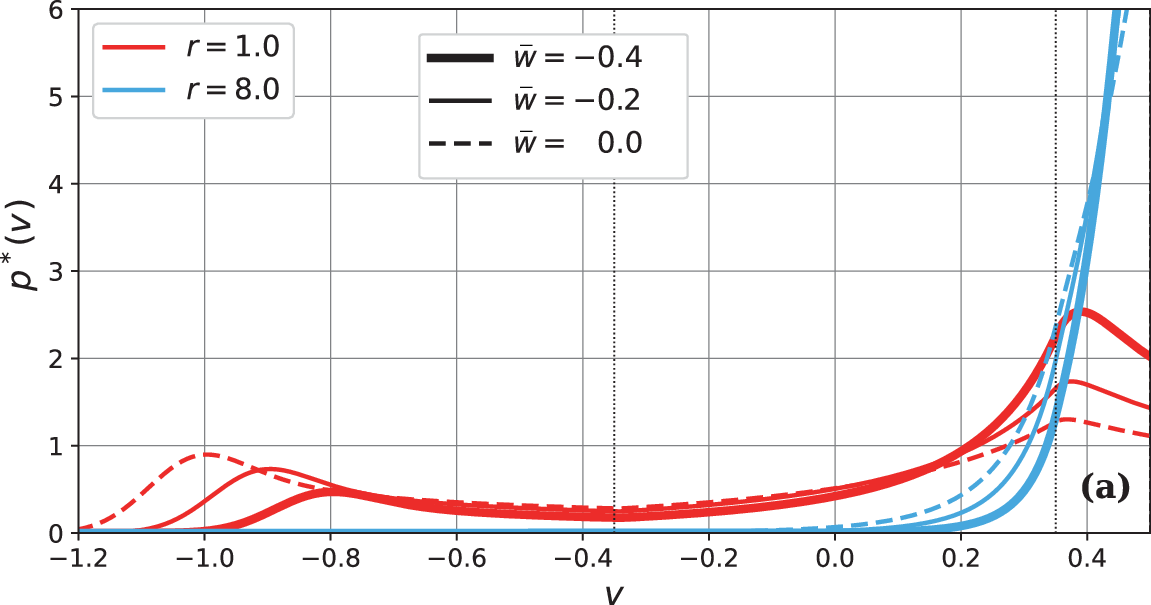}
               \includegraphics[width=0.9\linewidth]{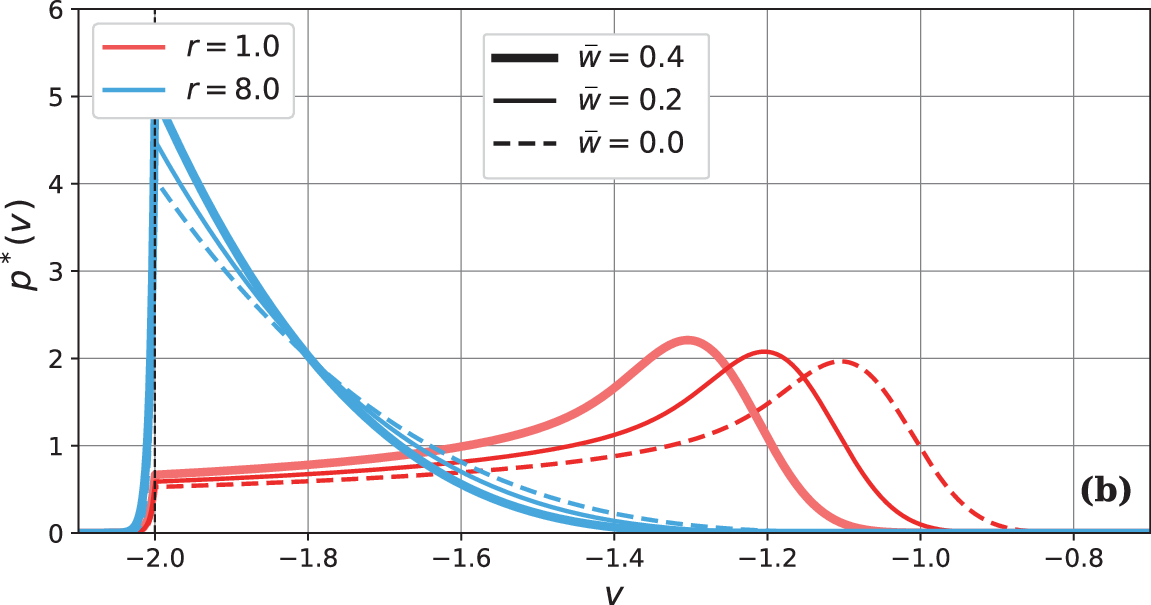}
                \includegraphics[width=0.9\linewidth]{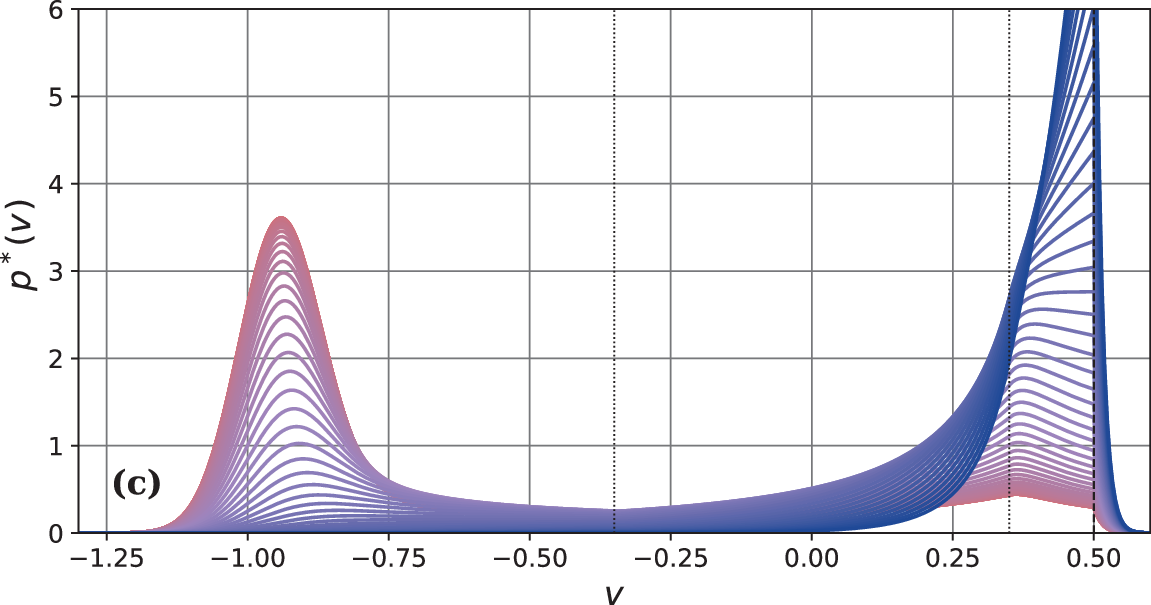}
           \caption{Diffusive analog of Figs. \ref{fig5}(b,c) with $D=0.01$. Plots of \(p^{*}(v)\) with resetting in the diffusive excitable regime for different values of the frozen slow variable $\overline w$ and the resetting rate $r$. Other parameters are $I_0=-1$, $a=0.35$, $\varepsilon =0.1$, $\gamma=0.5$, $m_{\pm }=-2$ and $m_0=1.14$. (a) $v_r=0.5$ (resetting to the right-hand branch) and (b) $v_r=-2$ (resetting to the left-hand branch). In both graphs, slower resetting allows for probability to accumulate near the fixed point on the fast nullcline. (c) Transition between stationary distributions over a range of resetting rates $r\in [0.2,8]$ for $v_r=0.5$ and $\overline w=-0.2$.}
\label{fig10}
\end{figure}

In each of the domains ${\mathcal I}_i$, $i\in \{+,-,0\}$, introduce the rescaled variable 
\begin{equation}
\label{calF}
z ={\mathcal F}_i(v):= \sqrt{\frac{|m_i|}{D}}\left(v+\frac{b_i+I_0-\overline w}{m_i}\right)
\end{equation}
 so that Eq. (\ref{u2}) can be rewritten in the piecewise form
\begin{align}
\frac{d^2u}{dz^2}
+
\left[
-\left(\frac{m_i}{2}+r\right)\frac{1}{|m_i|}
-\frac{z^2}{4}
\right]u(z)
=0,\quad z\in {\mathcal F}_i({\mathcal I}_i)
\end{align}
or, equivalently,
\begin{align}
\frac{d^2u}{dz^2}
+
\left[
\nu_i+\frac{1}{2}-\frac{z^2}{4}
\right]u(z)
=0,\quad z\in {\mathcal F}_i({\mathcal I}_i),
\label{eq:nu_nu}
\end{align}
where
\begin{equation}
\qquad
\nu_i:=
-
\left(\frac{m_i}{2}+r\right)\frac{1}{|m_i|}
-\frac{1}{2}.
\end{equation}
Eq. (\ref{eq:nu_nu} ) is equivalent to the Weber equation \cite{white2017boundstateenergiesusing}.
Hence, for generic values of \(\nu_i\), we obtain a solution
of the form
\[
u(z)=A_iD_{\nu_i}(z)+B_iD_{\nu_i}(-z),\quad z\in {\mathcal F}_i({\mathcal I}_i),
\]
where \(D_{\nu_i}(z)\), \(D_{\nu_i}(-z)\) are the parabolic cylinder functions.
(Note that in the non-generic case \(\nu_i\in\mathbb{N} \cup \{0\}\), \(D_{\nu_i}(z)\) and \(D_{\nu_i}(-z)\) are linearly dependent
and hence do not form a basis of the solution space \cite{dunster2021uniformasymptoticexpansionssolutions}.)  We conclude that the
piecewise stationary density away from the reset point $v=v_r$ and transition points $v=\pm a$ has the general form
\begin{equation}
    p^*(v)=e^{S_i(v)}\Bigl[A_iD_{\nu_i}(z)+B_iD_{\nu_i}(-z)\Bigr],\quad z\in {\mathcal F}_i({\mathcal I}_i),
    \label{eq:diff_sol}
\end{equation}
where \(i \in \{+, 0, -\} \).

At this point we note that, unlike the non-diffusive case, 
the system can move beyond restrictions imposed by
an otherwise deterministic flow, and so the support of $v$
is all of \(\mathbb{R}\). This means that we need to impose integrability at infinity in order to ensure unit normalization of the NESS.
The asymptotic behaviour of the parabolic cylinder functions as $z\to +\infty$ is
\(\nu\) is \cite{AbramowitzStegun1964}
\begin{subequations}
\begin{align}
   & D_\nu(z)\sim z^\nu e^{-z^2/4},
    \\
  & D_\nu(-z)\sim z^\nu e^{-z^2/4}
    -\frac{\sqrt{2\pi}}{\Gamma(-\nu)}\,z^{-\nu-1}e^{z^2/4}.
\end{align}
\end{subequations}
We see that as \(z\to +\infty\), \(D_\nu(z)\) is exponentially decaying whereas \(D_\nu(-z)\) is not.
Hence, in order to ensure integrability of the solution (\ref{eq:diff_sol}), we require \(A_-=B_+=0\). 
It remains to deal with the matching conditions (\ref{con1})-(\ref{con3}). For the sake of illustration, suppose that the reset point is located on the rightmost branch, that is \(v_r>a\). (Again, a similar analysis can be carried out if the reset point is located on one of the other branches.) It is also convenient to introduce the
modified basis functions
\begin{equation}
\phi_{i,\pm}(v)=e^{S_i(v)}D_{\nu_i}\bigl(\pm \calF_i(v)\bigr)
\end{equation}
and the corresponding fluxes
\begin{align}
J_{i,\pm}(v):=g(v)\phi_{i,\pm}(v)-D\phi_{i,\pm}'(v) .
\end{align}
Taking into account the integrability conditions and the jump discontinuity across $v_r$ we can then set
\begin{subequations}
\begin{align}
p^*(v )&=B_-\,\phi_{-,-}(v), \quad v<-a,\\
p^*(v )&=A_0\,\phi_{0,+}(v)+B_0\,\phi_{0,-}(v), \quad -a< v< a,\\
p^*(v )&=A_<\,\phi_{+,+}(v)+B_<\,\phi_{+,-}(v), \quad a< v<v_r,\\
p^*(v )&=A_>\,\phi_{+,+}(v), \quad v>v_r.
\end{align}
\end{subequations}
Substituting into Eqs. (\ref{con1})-(\ref{con3}) then yields the following matrix equation for the coefficients:

\begin{figure}[h!]
    \centering
    \includegraphics[width=\linewidth]{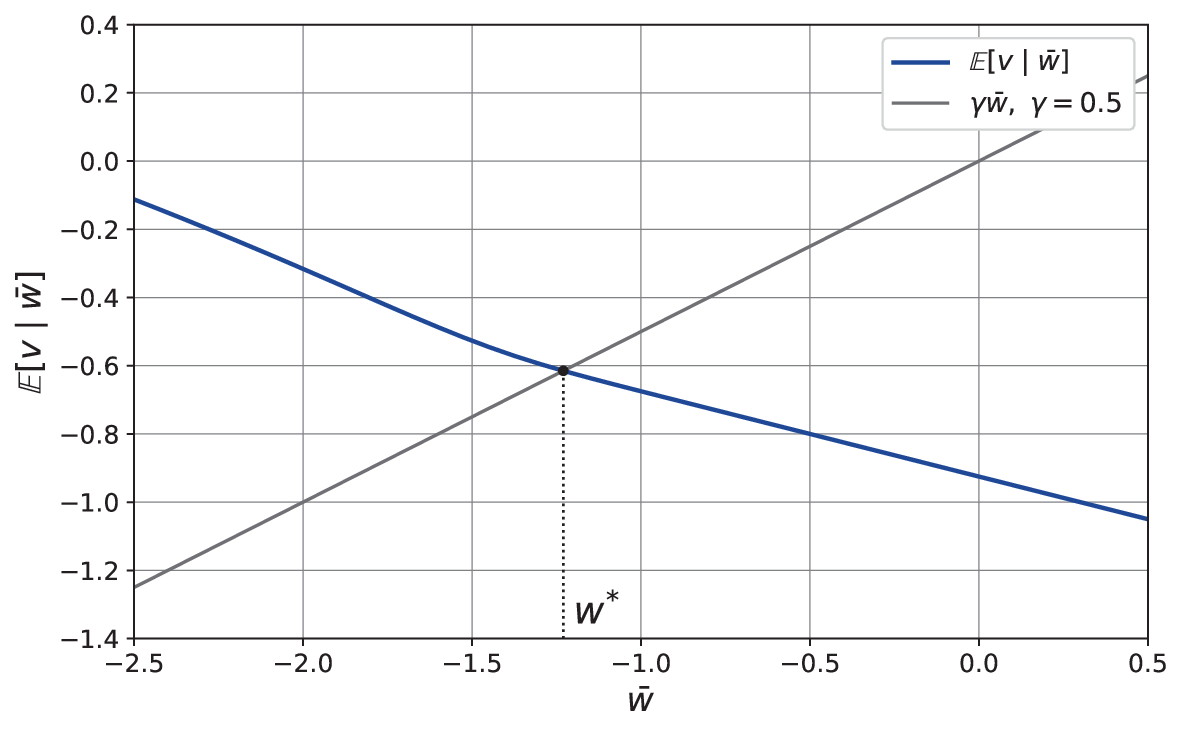}
    \caption{Numerically evaluated plot of \( \mathbb{E}[v|w] \) as a function of $w$ for the McKean model with stochastic resetting and diffusion. Same parameter values as Fig. \ref{fig9}(a,b). We use the corresponding NESS calculated in terms of parabolic cylinder functions.  The intersection with the straight line $y=\gamma w$ yields the fixed point \(w^* \approx -1.25\).}
    \label{fig11}
\end{figure}

\begin{figure}[h!]
    \centering
        \includegraphics[width=\linewidth]{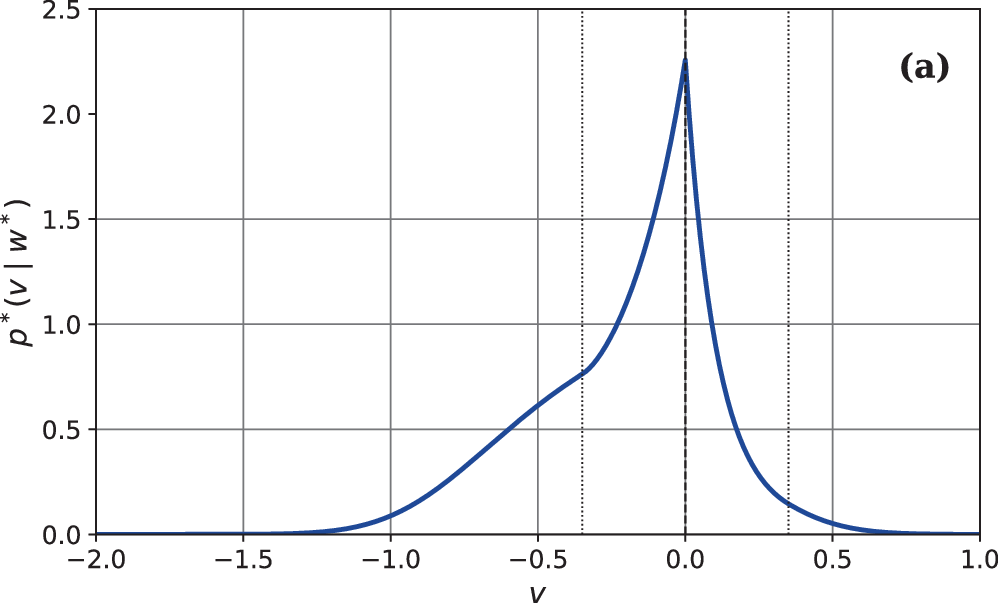}
               \includegraphics[width=\linewidth]{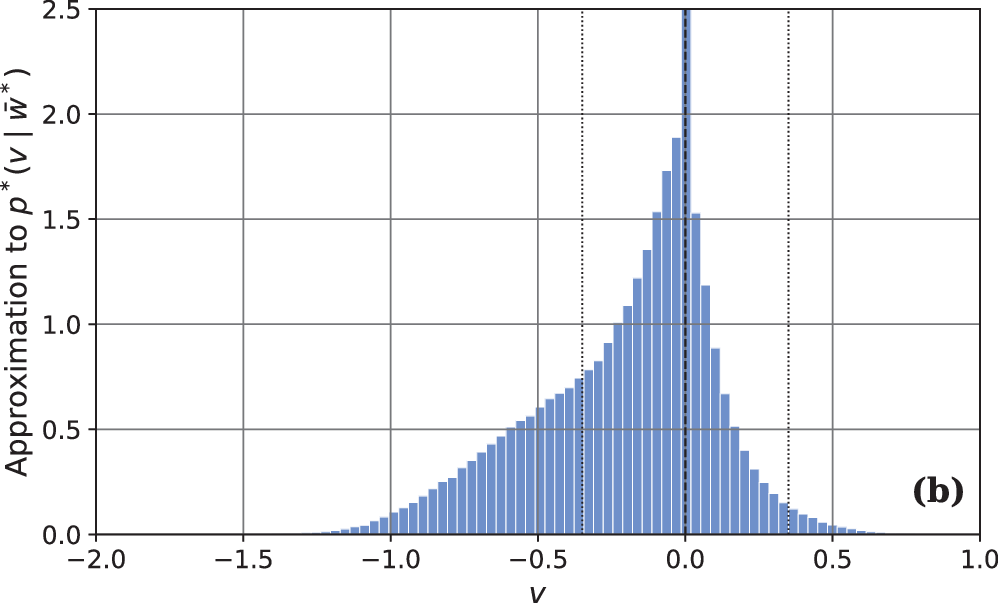}
            \caption{ (a) The NESS as determined by the parabolic cylinder functions for $r=3$, $v_r=0$ and $D=0.1$. Other parameters are $I_0=-1$, $a=0.35$, $\varepsilon =0.06$, $\gamma=0.5$, $m_{\pm }=-2$ and $m_0=1.14$. (b) Corresponding histogram obtained from numerical simulations of the full system. The fixed point of the slow variable is \(w^* \approx -0.419\).}
    \label{fig12}
\end{figure}

\begin{widetext}
\begin{equation}
\begin{pmatrix}
\phi_{-,-}(-a) & -\phi_{0,+}(-a) & -\phi_{0,-}(-a) & 0 & 0 & 0 \\
J_{-,-}(-a) & -J_{0,+}(-a) & -J_{0,-}(-a) & 0 & 0 & 0 \\
0 & \phi_{0,+}(a) & \phi_{0,-}(a) & -\phi_{+,+}(a) & -\phi_{+,-}(a) & 0 \\
0 & J_{0,+}(a) & J_{0,-}(a) & -J_{+,+}(a) & -J_{+,-}(a) & 0 \\
0 & 0 & 0 & \phi_{+,+}(v_r) & \phi_{+,-}(v_r) & -\phi_{+,+}(v_r) \\
0 & 0 & 0 & -J_{+,+}(v_r) & -J_{+,-}(v_r) & J_{+,+}(v_r)
\end{pmatrix}
\begin{pmatrix}
B_-\\
A_0\\
B_0\\
A_<\\
B_<\\
A_>
\end{pmatrix}
=
\begin{pmatrix}
0\\
0\\
0\\
0\\
0\\
r
\end{pmatrix}.
\end{equation}
\end{widetext}

\begin{figure}[t!]
    \centering    
        \includegraphics[width=\linewidth]{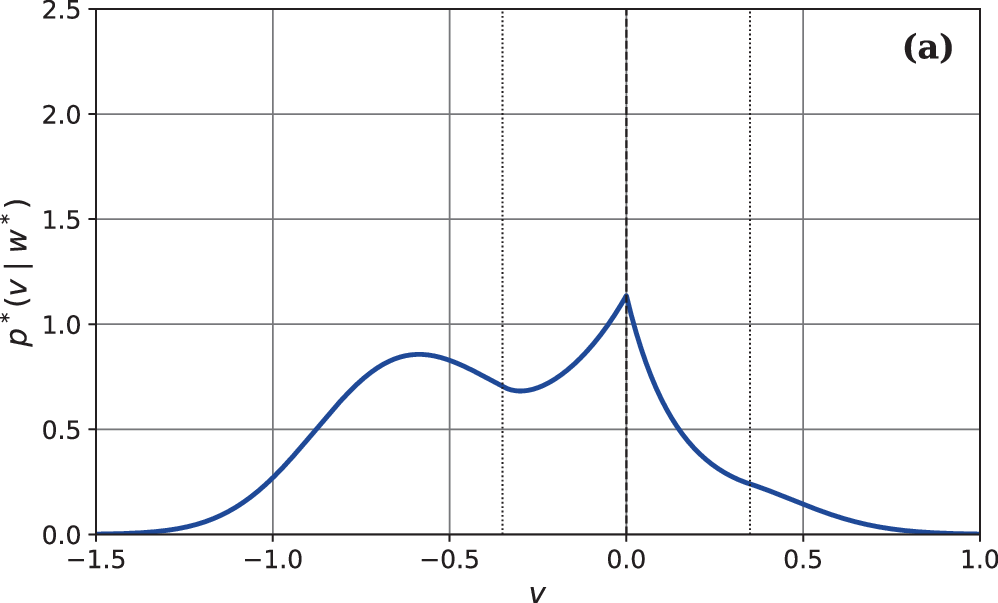}
        \includegraphics[width=\linewidth]{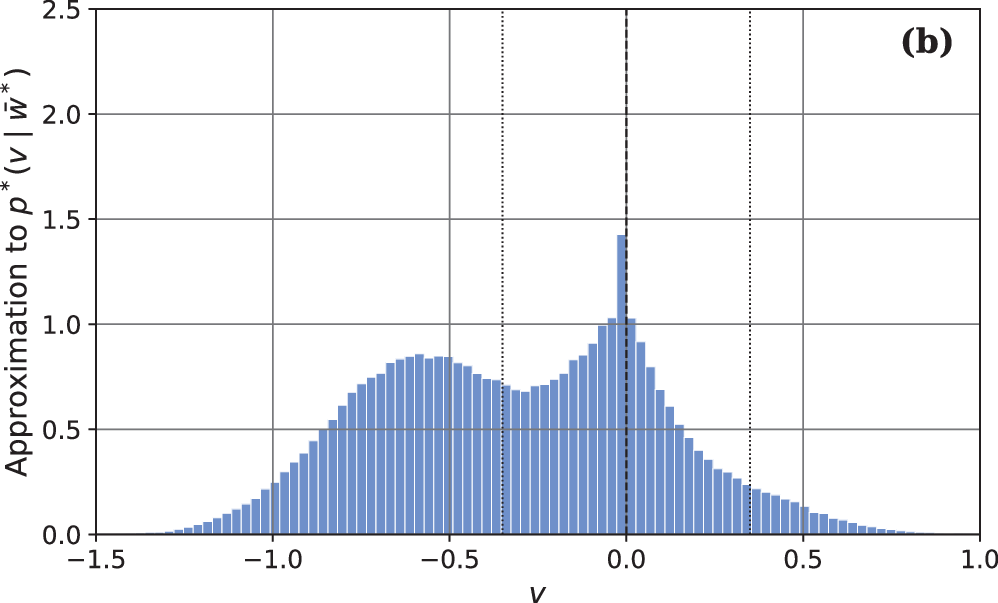}
    \caption{The same as Fig. \ref{fig12} with \(r = 1.0\) and \(D=0.1\). In this example $w^* \approx -0.585$.}
    \label{fig13}
\end{figure}

\begin{figure}[t!]
    \centering
            \includegraphics[width=\linewidth]{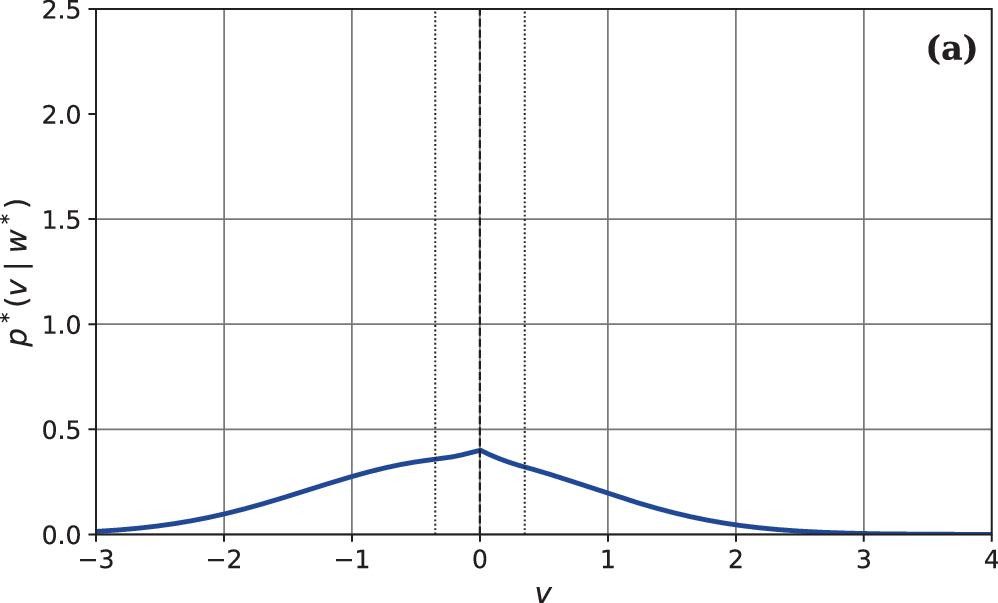}
        \includegraphics[width=\linewidth]{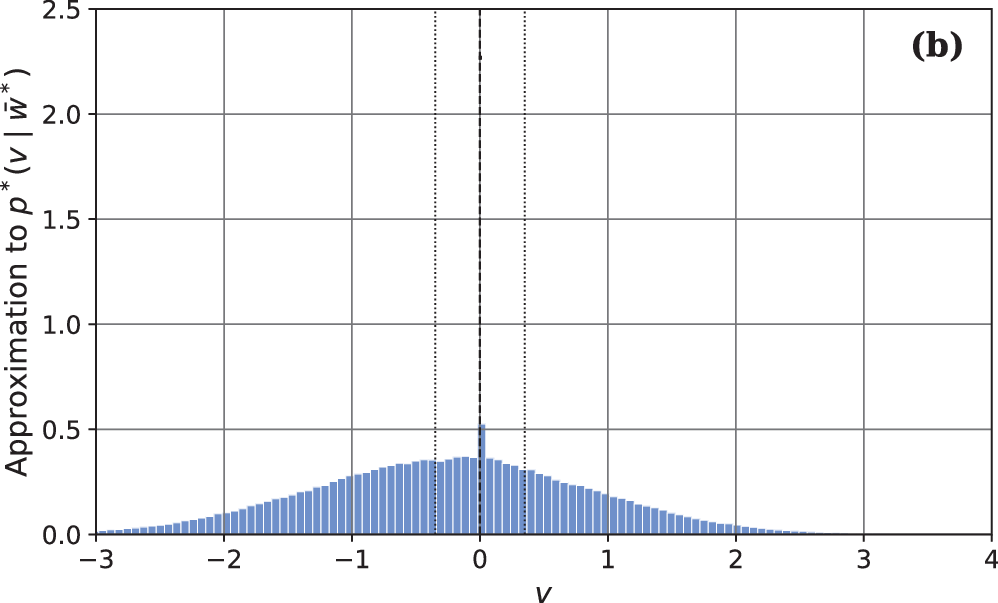}
    \caption{The same as Fig. \ref{fig12} with \(r = 1.0\) and \(D=2\). In this example $w^* \approx -0.518$.  The underlying deterministic structure vanishes in the case of large $D $ and the NESS becomes approximately Gaussian as expected for a pure diffusion process.}
    \label{fig14}
\end{figure}

Inverting the above inhomogeneous matrix equation determines the remaining coefficients, and we thus obtain the full solution to the layer problem in the diffusive case. In practice, the matrix version is carried out numerically. In Fig. \ref{fig10} we show plots of the NESS for the same examples considered in Fig. \ref{fig5}(b,c) except that now $D>0$. Whilst the support in both cases has become \(\mathbb{R}\) under the diffusion, the difference in qualitative behaviour comes as a consequence of the difference in deterministic dynamics at different points along the phase plane. Although diffusion removes the fixed points in a strict deterministic sense, the underlying deterministic dynamics still cause probability to accumulate around these points.

Having obtained the NESS of the layer problem, we can now determine the slow dynamics according to Eqs. (\ref{slow}) and (\ref{sfav}), and thus check that the numerically obtained equilibrium $w^*\approx -1.25$ in Fig. \ref{fig9}(b) is a root of Eq. (\ref{FPw}). This is confirmed in Fig. \ref{fig11} where we plot ${\mathbb E}[V|w]$ as a function of $w$ for the same parameter values as Fig. \ref{fig9}(b). Further validation of the slow-fast analysis can be obtained by comparing the analytically calculated stationary NESS $p^*(v|w^*)$ with numerically generated histograms based on simulations of the full system. Examples are shown in Figs. \ref{fig12}, \ref{fig13} and \ref{fig14}, where the effect of varying diffusivity and reset rate is evident. As the noise level increases, the concentration of probability near the equilibrium and reset locations becomes less pronounced, as the underlying deterministic dynamics now has a lesser effect. It can be seen that there is excellent agreement between the histograms and analytic density profiles.

\subsection{Numerical comparison with the FitzHugh--Nagumo system}

\begin{figure}[b!]
    \centering
        \includegraphics[width=\linewidth]{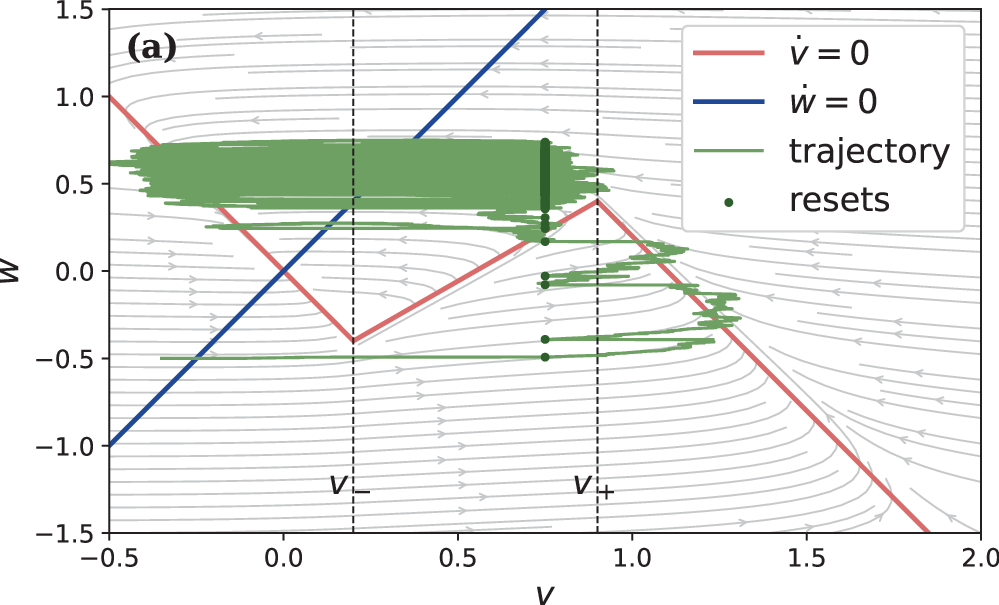}
          \includegraphics[width=\linewidth]{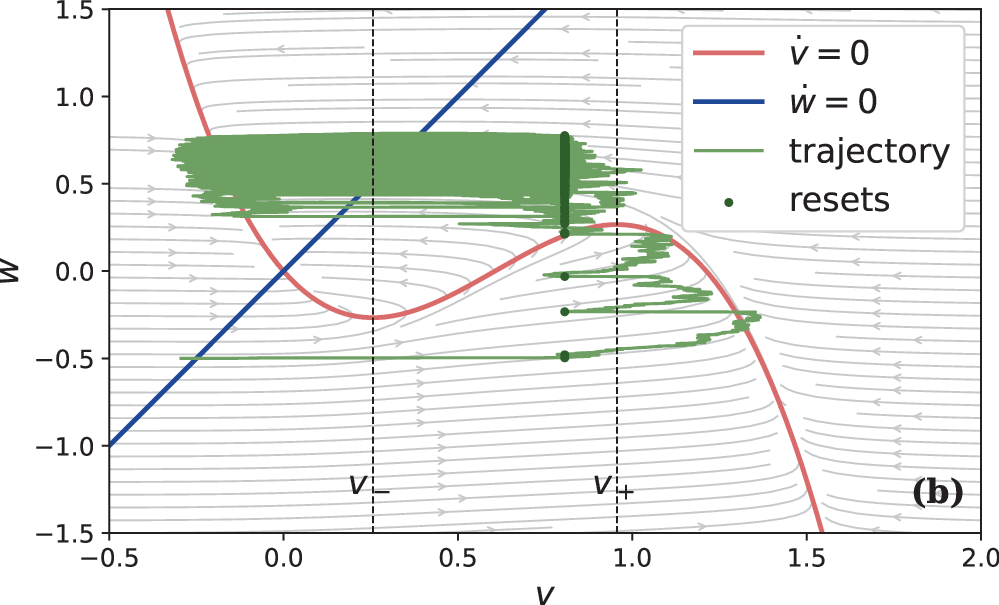}
       \caption{Phase-plane plots in the excitable regime with diffusion and stochastic resetting for (a) the McKean model and (b) the FN model. The fast nullclines are shown in red and the slow nullclines are shown in blue. In both cases the reset position lies on the middle branch. For the McKean model we consider a shifted version of Fig. \ref{fig2}(a) with turning points at $v_-=0.2$ and $v_+=0.9$; the reset state is $v_r=0.75$. For the FN model the cubic nonlinearity is chosen so that the turning points occur at $v_-\approx 0.2562$ and $v_+\approx 0.9562$, and the gradient at the centre of the middle branch agrees with the McKean model.; the reset point is now $v_r=0.806$. The remaining parameters are \(\varepsilon=0.1\), \(\gamma=0.5\), \(D=0.01\) and \(r=1.0\).}
    \label{fig15}
\end{figure}

\begin{figure}[t!]
    \centering
    \includegraphics[width=\linewidth]{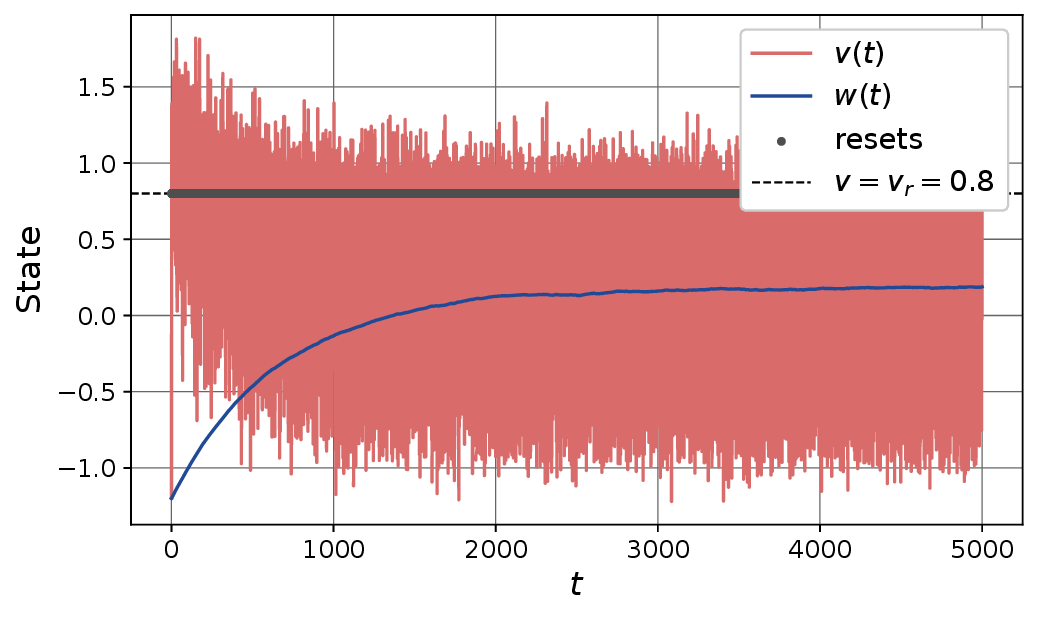}
    \caption{Numerical time series showing the evolution of the fast variable \(v(t)\) and slow variable \(w(t)\) in the FN model with diffusion and resetting. Same parameters as Fig. \ref{fig15}(b). The slow variable converges to a fixed point after approximately 3000 time steps, while the fast variable continues to fluctuate due to diffusion and resetting.}
    \label{fig16}
\end{figure}

\begin{figure}[h!]
    \centering
        \includegraphics[width=\linewidth]{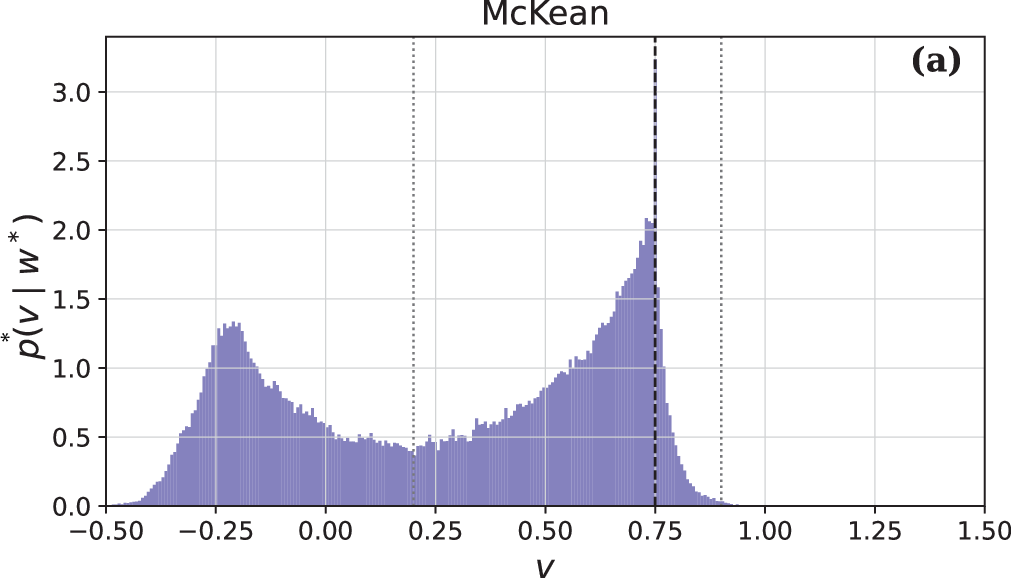}
         \includegraphics[width=\linewidth]{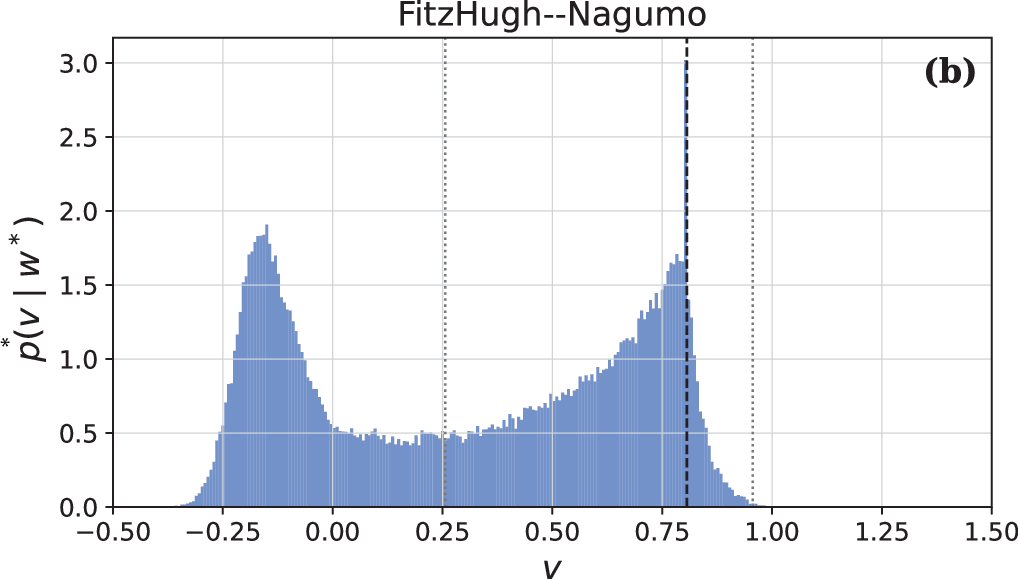}
    \caption{NESS \(p^*(v| w^*)\) with diffusion and resetting, computed from long-time simulations for (a) the McKean model and (b) the FN model. Same parameters as Figs. \ref{fig15}(a) and \ref{fig15}(b), respectively.}
    \label{fig17}
\end{figure}

We recall that the piecewise linear McKean model was chosen as an analytically tractable substitute for the FN equations in the case with diffusion. Extensive numerical simulations shows that both models exhibit similar behavior in the \(\varepsilon \to 0^+\) limit. This is illustrated in Fig. \ref{fig15}, which compares the diffusive FN and McKean models with fast resetting to the middle branch of the $v$-nullcline. For a fair comparison, we take the turning points of both models to be similar and take the gradient at the centre of the middle branch (\(v=0\)) to be the same. In both models the variable $w(t)$ slowly increases, resulting in a sudden shift in the center-of-mass of the quasi-stationary density for $v$ and eventual convergence to a fixed point of the slow dynamics. Further confirmation of the convergence of $w(t)$ to an equilibrium is shown in Fig. \ref{fig16}. Finally, in Fig. \ref{fig17} we show histograms of the equilibrium NESS $p^*(v|w^*)$ for both models, which are similar both qualitatively and quantitatively.

\setcounter{equation}{0}

\section{Asymptotic analysis of the slow diffusion limit}

In this section we show how the NESS of the non-diffusive case, which is the solution of Eq. (\ref{eq:stat_D0}), can be recovered by taking the limit $D\rightarrow 0^+$ of the general solution (\ref{eq:diff_sol}). This requires an asymptotic analysis of the parabolic cylinder functions. Following Ref. \onlinecite{AbramowitzStegun1964}, we define
\begin{subequations}
\begin{align}
U(a,z) &= D_{-a-\frac12}(z),\\
V(a,z) &= \frac{1}{\pi}\Gamma\!\left(\frac12+a\right)
\left[
\sin(\pi a)\,D_{-a-\frac12}(z) + D_{-a-\frac12}(-z)
\right],
\end{align}
\end{subequations}
where $\Gamma(1/2+a)$ is the gamma function.
Setting $\nu=-a-\frac12$ and noting that
\[
\sin(\pi a)=\sin\!\left(-\nu\pi-\frac12\pi\right)=-\cos(\pi\nu),
\]
we may rewrite these as
\begin{subequations}
\begin{align}
U\!\left(-\nu-\frac12,z\right) &= D_\nu(z),\\
V\!\left(-\nu-\frac12,z\right) &= \frac{\Gamma(-\nu)}{\pi}
\left[
\cos(\pi\nu)D_\nu(z) + D_\nu(-z)
\right].
\label{eq:asymp_2}
\end{align}
\end{subequations}
Recalling that the rescaled variable \(z\) is defined according to Eq. (\ref{calF}),
the presence of the diffusivity \(D\) in the denominator means that a small-diffusion asymptotic analysis is equivalent to considering the
asymptotic behaviour of the parabolic cylinder functions for large $|z|$.
From Ref. \onlinecite{AbramowitzStegun1964}, we have that the large-$z$ asymptotics of the alternative solution basis $U$ and $V$ are given by
\begin{subequations}
\begin{align}
U(a,z) &\sim e^{-\frac14 z^2}z^{-a-\frac12}
\left[
1-\frac{(a+\frac12)(a+\frac32)}{2z^2}+O(z^{-4})
\right],\\
V(a,z) &\sim \sqrt{\frac{2}{\pi}}\,e^{\frac14 z^2}z^{a-\frac12}
\left[
1+\frac{(a-\frac12)(a-\frac32)}{2z^2}+O(z^{-4})
\right].
\end{align}
\end{subequations}
Rearranging Eq. (\ref{eq:asymp_2}) and again writing $a=-\nu-\frac12$, the large-$z$
asymptotics of the chosen solution basis now become
\begin{align}
    D_\nu(z)&=U\!\left(-\nu-\frac12,z\right)\nonumber \\
&\sim e^{-\frac14 z^2}z^\nu
\left(
1-\frac{\nu(\nu-1)}{2z^2}+O(z^{-4})
\right)
\label{eq:final_asymp_2}
\end{align}
for the first solution. Then, by rewriting the second parabolic cylinder function in terms
of the first as
\begin{equation}
    D_\nu(-z)=\cos(\pi\nu)D_\nu(z)+\frac{\pi}{\Gamma(-\nu)}
V\!\left(-\nu-\frac12,z\right),
\end{equation}
its asymptotic behaviour becomes
\begin{align}
\label{eq:final_asymp_1}
&D_\nu(-z)\\
&\sim \cos(\pi\nu)e^{-\frac14 z^2}z^\nu
\left(
1-\frac{\nu(\nu-1)}{2z^2}+O(z^{-4})
\right) \nonumber \\ 
&\quad
+\frac{\sqrt{2\pi}}{\Gamma(-\nu)}e^{\frac14 z^2}z^{-\nu-1}
\left(
1+\frac{(\nu+1)(\nu+2)}{2z^2}+O(z^{-4})
\right).\nonumber
\end{align}

\begin{widetext}
Substituting the asymptotic expansions (\ref{eq:final_asymp_2}) and (\ref{eq:final_asymp_1}) into the general solution (\ref{eq:diff_sol}) yields the following
expansion up to second order in $z=\calF_i(v)$:
\begin{align}
\nonumber
p_i(v;D)
&\sim e^{S_i(v)}
\Bigg[
\bigl(A_i+B_i\cos(\pi\nu_i)\bigr)e^{-\calF_i(v)^2/4}\calF_i(v)^{\nu_i}
\left(
1-\frac{\nu_i(\nu_i-1)}{2\calF_i(v)^2}+O\!\bigl(\calF_i(v)^{-4}\bigr)
\right)\\[16pt]
&\qquad\qquad
+ B_i\frac{\sqrt{2\pi}}{\Gamma(-\nu_i)}e^{\calF_i(v)^2/4}\calF_i(v)^{-\nu_i-1}
\left(
1+\frac{(\nu_i+1)(\nu_i+2)}{2\calF_i(v)^2}+O\!\bigl(\calF_i(v)^{-4}\bigr)
\right)
\Bigg].
\label{boo}
\end{align}
\end{widetext}
We now observe that the exponential factor 
on the first line of the right-hand side is governed by the exponent
\begin{equation}
\label{alph}
\alpha_i(v):=S_i(v)-\frac{\calF_i(v)^2}{4} = \frac{m_iv^2}{4D}+\frac{d_iv}{2D}
-\frac14\left[\frac{m_iv+d_i}{\sqrt{D|m_i|}}\right]^2,
\end{equation}
where $d_i=b_i+I_0-\overline w$,
while the exponential factor on the second line is governed by the exponent
\begin{equation}
\label{beta}
\beta_i(v):=S_i(v)+\frac{\calF_i(v)^2}{4}
=
\frac{m_iv^2}{4D}+\frac{d_iv}{2D}
+\frac14\left[\frac{m_iv+d_i}{\sqrt{D|m_i|}}\right]^2.
\end{equation}
On the outer branches we have $m_{\pm}<0$, so that
\begin{align}
\alpha_{\pm}(v)
&=
\frac{m_\pm v^2}{2D}+\frac{d_\pm v}{D}+\frac{d_\pm^2}{4Dm_\pm} 
\end{align}
and
\begin{align}
\beta_{\pm}(v)
=
-\frac{d_\pm^2}{4Dm_\pm}.
\end{align}
We note that the quadratic function $\alpha_{\pm}(v)$ has a maximum at $v_{\pm} = -d_{\pm}/m_{\pm}$ with $\alpha(v_{\pm})=-d_{\pm}^2/4Dm_{\pm}$. Hence, $\alpha_{\pm}(v) <\beta_i(v)$ for all $v\neq v_{\pm}$ so that the first line on the right-hand side of Eq. (\ref{boo}) is exponentially small compared to the second line in the small diffusion limit. We thus have the asymptotic solution 
\begin{align}
    p_\pm(v;D)\sim
B_\pm\frac{\sqrt{2\pi}}{\Gamma(-\nu_\pm)}
e^{-d_\pm^2/4Dm_\pm}
\calF_\pm(v)^{-\nu_\pm-1}\nonumber \\
\left[
1+\frac{(\nu_\pm+1)(\nu_\pm+2)}{2z_\pm(v)^2}
+O\!\bigl(\calF_\pm(v)^{-4}\bigr)
\right].
\label{eq:before}
\end{align}
Finally, from Eq. (\ref{calF}), we find that to leading order 
\begin{align}
p_\pm(v;D)&\propto (m_\pm v+d_\pm)^{-\nu_\pm-1}\nonumber \\
&=
\frac{|m_\pm v+d_\pm|^{-r/m_\pm}}{(m_\pm v+d_\pm)}.
\end{align}
We have used the fact that on the outer branches, the general definition of $\nu_\pm$ given by (\ref{eq:nu_nu}), becomes,
\begin{align}
    \nu_\pm &= -
\left(\frac{m_i}{2}+r\right)\frac{1}{|m_i|}
-\frac{1}{2} \nonumber \\
&=
\left(\frac{m_i}{2}+r\right)\frac{1}{m_i}
-\frac{1}{2} = \frac{r}{m_\pm}.
\end{align}
Thus, on the outer branches, the leading-order asymptotics coincide with the non-diffusive general solution (\ref {solpstar}) for $i=\pm $.

Note that it is also possible to recover the non-diffusive solution on the middle branch. In particular we have
\[
\alpha_0(v)
=
-\frac{d_0^2}{4Dm_0}
\]
and
\[
\beta_0(v)
=
\frac{m_0v^2}{2D}+\frac{d_0v}{D}+\frac{d_0^2}{4Dm_0}.
\]
Here the first term contributes only a global prefactor, whereas the
second term leads to a divergent contribution in the limit $|v\rightarrow \infty$. In this case we eliminate the resulting divergent term by setting $B_0=0$ such that
\[
p_0(v;D)\propto \frac{|m_0v+d_0|^{-r/m_0}}{m_0v+d_0}.
\]

\setcounter{equation}{0}

\section{Slow resetting regime $r=O(\varepsilon)$}

\begin{figure}[b!]
    \centering
    \includegraphics[width=\linewidth]{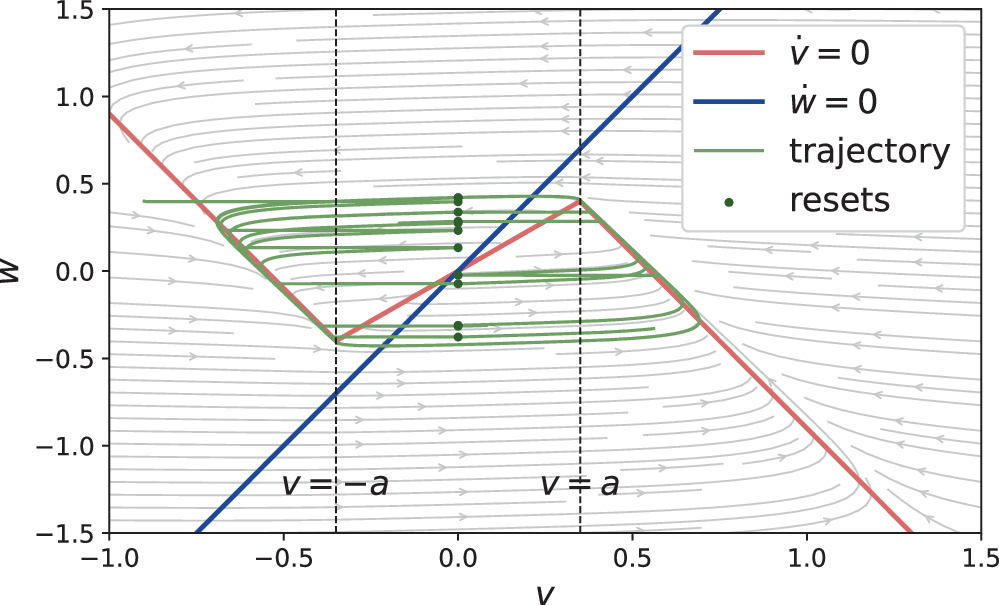}
    \caption{Phase-plane plot of a trajectory in the slow resetting, non-diffusive  oscillatory regime, with \(r = 0.05\),  $\varepsilon = 0.1$, $\gamma=0.5$, \(I_0 = 0.0\), $a=0.35$, $m_{\pm }=-2$, $m_0=1.14$ and \(v_r = 0.0\).}
    \label{fig18}
\end{figure}

So far we have focused on the fast-resetting regime, that is \(r = O(1)\). In this case the slow variable converges to an equilibrium irrespective of whether the underlying deterministic system operates in an excitable or oscillatory regime. That is, there are no persistent large excursions in the phase-plane. If instead \(r = O(\varepsilon)\), then resets occur on the slow timescale and a noisy version of the oscillatory regime re-emerges as illustrated in Fig. \ref{fig18}. In this section we analyze such behavior, focussing on the non-diffusive case ($D=0$). We briefly discuss the diffusive case at the end of the section.

\begin{figure}[t!]
    \centering
    \includegraphics[width=\linewidth]{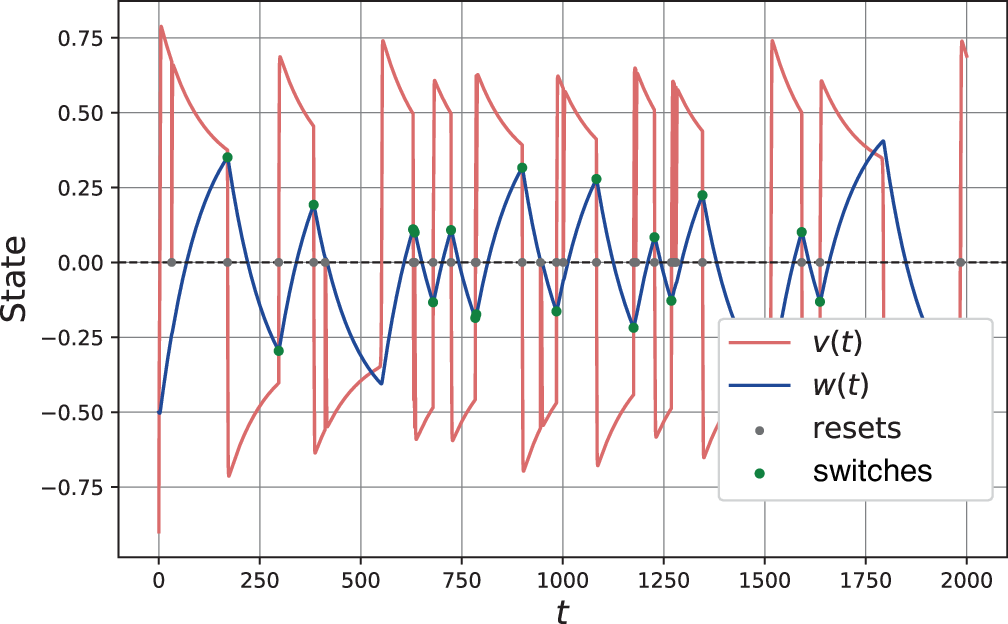}
    \caption{A time series, in the slow resetting oscillatory regime, depicting the slow and fast variable trajectories. At reset points the slow variable may change direction as a consequence of being transported to the opposite $v$-nullcline branch. Same parameters as Fig. \ref{fig18}.}
    \label{fig19}
\end{figure}

Since resets are now rare events in the limit \(\varepsilon\to 0^+\), the system spends most of the time moving along the $v$-nullcline according to the reduced problem (\ref{sf}).
Hence, between resets, the slow dynamics is determined by which of the two outer branches the system is currently on. That is, on the slow timescale $\tau=\varepsilon t$
\begin{equation}
    \frac{dw}{d\tau}=v_{\pm}(w)-\gamma w ,  
\end{equation}
where
\begin{equation}
    v_-(w)=\frac{w-I_0-b_-}{m_-}, \qquad
v_+(w)=\frac{w-I_0-b_+}{m_+}
\end{equation}
on the left-hand and right-hand branches, respectively. Suppose that the reset point lies somewhere on the middle branch as in Fig. \ref{fig18}. Recall that resetting acts only on the fast variable $v$ so that after each reset event, the system will rapidly return from $v_r$ to one of the outer branches with high probability. Hence, the only effect of resetting on the slow dynamics is to induce an instantaneous switch from one outer branch to its opposite. However, this will only occur if resetting crosses the middle branch. Introduce the critical threshold
\[
w_r = f(v_r)+I_0,
\]
such that $(v_r,w_r)$ lies on the middle branch. It follows that the slow dynamics will switch from the right-hand branch to the left-hand branch if $w>w_r$ and vice versa when $w < w_r$. This type of resetting-induced switching is illustrated by the time series plots in Fig. \ref{fig19}. Since resetting is a stochastic process, it follows that switching of the slow dynamics is also stochastic and we can construct a corresponding probabilistic model.

First, we define the pair of probability densities $p_{\pm}(w,\tau)$ where \(p_+\) corresponds to the right branch and \(p_-\) to the left. Since the reduced dynamics evolves along the attracting outer branches, the
relevant \(w\)-interval is bounded by the extremal values of the $v$-nullcline, namely, 
$w_{\pm }:= f(\pm a)+I_0$.
Accordingly, the stationary densities are supported on \(w\in[w_-,w_+]\), and leads to
the normalisation condition 
\begin{equation}
\int_{w_-}^{w_+}\bigg [p_+(w,\tau)+p_-(w,\tau)\bigg ]dw=1.
\label{norm}
\end{equation}
It is clear that probability changes only through deterministic transport along a
branch between switching events, together with loss and gain terms due to the subset of reset events that switch trajectories to a
different branch. Hence, we have the generalized Liouville equations
\begin{widetext}
\begin{subequations}
\begin{equation}
    \frac{\partial p_+}{\partial \tau}(w,\tau)
=
-\frac{\partial}{\partial w}
\Bigl[ \bigl(v_+(w)-\gamma w\bigr)p_+(w,\tau)\Bigr]
-rp_+(w,\tau)
+r\mathbf{1}_{\{w<w_r\}}\bigl(p_+(w,\tau)+p_-(w,\tau)\bigr),
\end{equation}
\begin{equation}
    \frac{\partial p_-}{\partial \tau}(w,\tau)
=
-\frac{\partial}{\partial w}
\Bigl[\bigl(v_-(w)-\gamma w\bigr)p_-(w,\tau)\Bigr]
-rp_-(w,\tau)
+r\mathbf{1}_{\{w>w_r\}}\bigl(p_+(w,\tau)+p_-(w,\tau)\bigr),
\end{equation}
\end{subequations}
\end{widetext}
where \(\mathbf{1}_{\{A\}}\) is the indicator function on the event \(A\). We have also rescaled the resetting rate according to $r\rightarrow \varepsilon r$.

We now look for a stationary solution by setting all time derivatives to zero. For
\(w<w_r\), by the definition of the indicator function,
\[
\frac{d}{dw}\Bigl[\bigl(v_+(w)-\gamma w\bigr)p_+(w)\Bigr]
=
r p_-(w),
\]
\[
\frac{d}{dw}\Bigl[\bigl(v_-(w)-\gamma w\bigr)p_-(w)\Bigr]
=
-r p_-(w),
\]
while for \(w>w_r\),
\[
\frac{d}{dw}\Bigl[\bigl(v_+(w)-\gamma w\bigr)p_+(w)\Bigr]
=
-r p_+(w),
\]
\[
\frac{d}{dw}\Bigl[\bigl(v_-(w)-\gamma w\bigr)p_-(w)\Bigr]
=
r p_+(w).
\]
These equations describe changes in branch occupation purely as a consequence of
reset-induced gains and losses. They are standard first-order ODEs. For
\(w<w_r\), the second equation yields
\begin{equation}
p_-(w)
=
\frac{C_1}{v_-(w)-\gamma w}
\exp\left(
-\int^w \frac{r}{v_-(s)-\gamma s}\,ds
\right).
\label{eq:pm-below-w0}
\end{equation}
Substituting this into the first equation then gives,
\begin{align}
  &  \frac{d}{dw}\Bigl[\bigl(v_+(w)-\gamma w\bigr)p_+(w)\Bigr]\\
&=
\frac{rC_1}{v_-(w)-\gamma w}
\exp\left(
-\int^w \frac{r}{v_-(s)-\gamma s}\,ds
\right).\nonumber 
\end{align}
Now note that
\begin{align}
 &   \frac{r}{v_-(w)-\gamma w}
\exp\left(
-\int^w \frac{r}{v_-(s)-\gamma s}\,ds
\right) \nonumber \\
&=
-\frac{d}{dw}
\exp\left(
-\int^w \frac{r}{v_-(s)-\gamma s}\,ds
\right).
\end{align}
Hence, after integrating, we obtain
\begin{equation}
    p_+(w)
=
\frac{
C_2 - C_1\exp\left(
-\int^w \frac{r}{v_-(s)-\gamma s}\,ds
\right)
}{
v_+(w)-\gamma w
}.
\end{equation}
An identical argument for \(w>w_r\) gives
\begin{equation}
    p_+(w)
=
\frac{C_3}{v_+(w)-\gamma w}
\exp\left(
-\int^w \frac{r}{v_+(s)-\gamma s}\,ds
\right),
\end{equation}
and
\begin{equation}
    p_-(w)
=
\frac{
C_4 - C_3\exp\left(
-\int^w \frac{r}{v_+(s)-\gamma s}\,ds
\right)
}{
v_-(w)-\gamma w
}.
\end{equation}
Finally, recall that
\[
v_-(w)-\gamma w
=
\left(\frac{1}{m_-}-\gamma\right)w-\frac{I_0+b_-}{m_-},
\]
\[
v_+(w)-\gamma w
=
\left(\frac{1}{m_+}-\gamma\right)w-\frac{I_0+b_+}{m_+}.
\]
Hence the integrals appearing in the stationary densities can be evaluated in
closed form:
\begin{align}
   & \int \frac{r}{v_i(w)-\gamma w}\,dw
   \\
   & =
    \frac{r}{\left(\frac{1}{m_i}-\gamma\right)}
    \log\left|
    \left(\frac{1}{m_i}-\gamma\right)w-\frac{I_0+b_i}{m_i}
    \right|. \nonumber
\end{align}

To determine \(C_1,C_2,C_3,C_4\), we first define the branch-wise slow fluxes
\begin{equation}
    J_{\pm}(w):=\bigl(v_{\pm}(w)-\gamma w\bigr)p_{\pm}(w).
    \end{equation}
Since resetting remains confined to the fast variable, there is no point source
in \(w\). Consequently, these fluxes must be continuous at the reset
threshold \(w=w_r\). This yields
\[
J_+(w_r^-)=J_+(w_r^+),\qquad J_-(w_r^-)=J_-(w_r^+),
\]
that is,
\begin{align}
  &  C_2 - C_1\exp\left(
-\int^{w_r}\frac{r}{v_-(s)-\gamma s}\,ds
\right) \nonumber \\
&=
C_3\exp\left(
-\int^{w_r}\frac{r}{v_+(s)-\gamma s}\,ds
\right),
\end{align}
and
\begin{align}
   & C_1\exp\left(
-\int^{w_r}\frac{r}{v_-(s)-\gamma s}\,ds
\right)
\nonumber \\
&=
C_4 - C_3\exp\left(
-\int^{w_r}\frac{r}{v_+(s)-\gamma s}\,ds
\right).
\end{align}
Adding these two equations shows immediately that \(C_2=C_4\). 

\begin{figure}[t!]
    \centering
    \includegraphics[width=\linewidth]{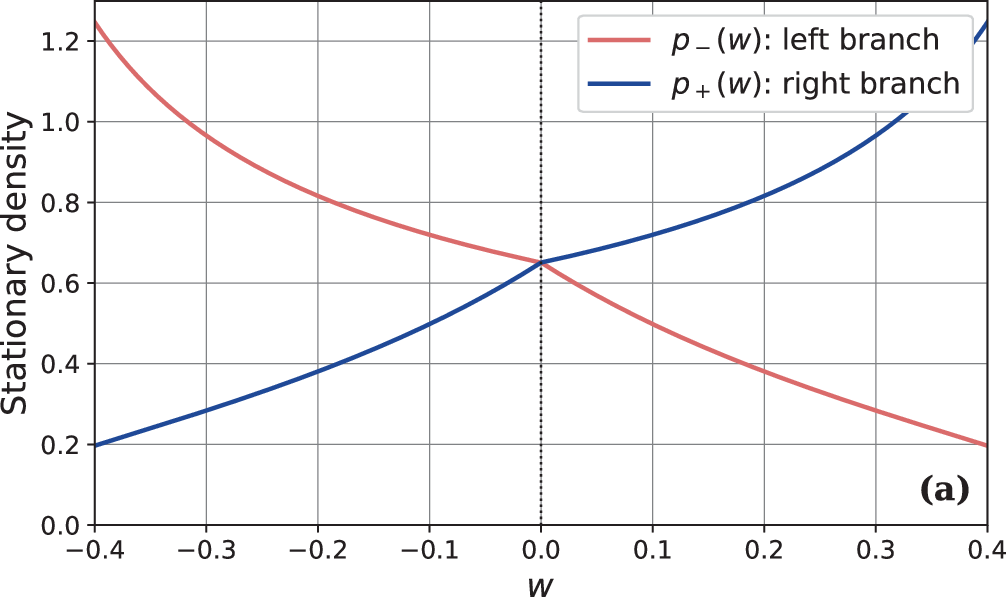}
    \includegraphics[width=\linewidth]{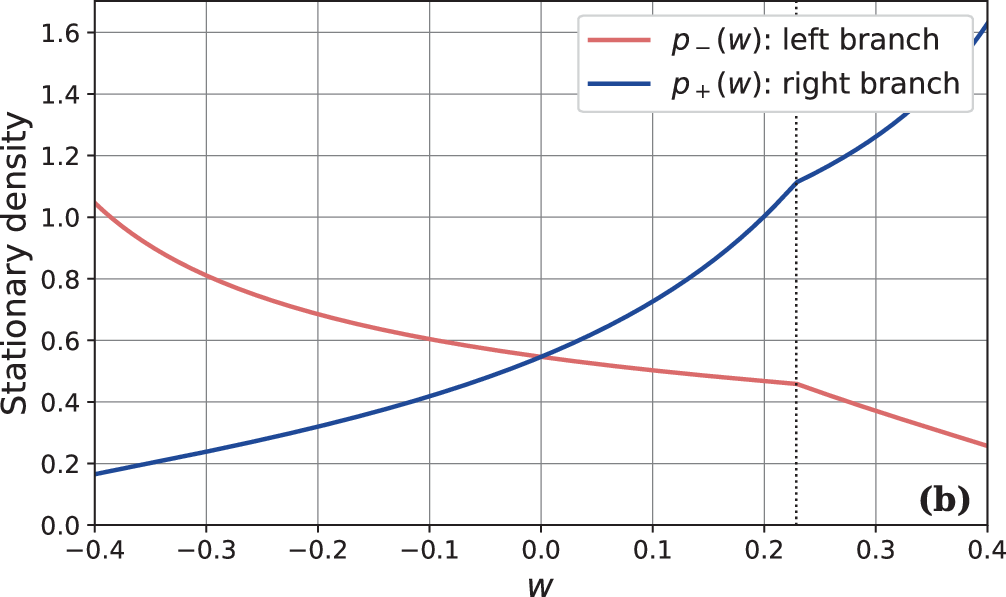}
    \caption{Analytical stationary branch densities in the non-diffusive
slow-resetting oscillatory regime. The red curve shows $p_-(w)$ on
the left attracting branch and the blue curve shows $p_+(w)$ on the
right attracting branch. Panel (a) shows the symmetric case
$v_r=0$, for which $w_r=0$. Panel (b) shows the asymmetric case
$v_r=0.2$, for which $w_r\approx0.2286$. The dotted vertical lines
mark the reset threshold $w=w_r$. Other parameters are $D=0$,
$r=0.05$, $\varepsilon=0.1$, $\gamma=0.5$, $I_0=0$, $a=0.35$,
$m_\pm=-2$ and $m_0=1.14$.}
    \label{fig20}
\end{figure}

The remaining conditions arise at the local minimum and maximum values, \(w_-\) and \(w_+\). To justify
these, note that the reduced slow description is only valid while trajectories
evolve along the attracting outer branches. Once a trajectory reaches a local extremum,
this reduced description breaks down and the full fast--slow system must again be
considered. In particular, the trajectory then undergoes a fast jump from one
attracting branch to the other. Crucially, during such a jump the slow variable is unchanged to leading order.
Hence, in
the singular limit \(\varepsilon\to0^+\), the fast transition between branches occurs
at fixed \(w\). Moreover, this jump is deterministic and carries no creation or
loss of probability mass. Therefore, conservation of probability implies that
the outgoing slow flux from one branch must equal the incoming slow flux onto the
other. Thus, at the right extreme point \(w=w_+\), where mass leaves the right attracting branch
and is transferred to the left attracting branch, conservation of slow
flux gives
\[
J_+(w_+) + J_-(w_+) = 0.
\]
Similarly, at the left local extreme point \(w=w_-\), where mass leaves the left attracting
branch and is transferred to the right attracting branch, we impose
\[
J_-(w_-) + J_+(w_-) = 0.
\]
Written out explicitly, these conditions are
\begin{equation}
    \bigl(v_+(w_+)-\gamma w_+\bigr)p_+(w_+)
+
\bigl(v_-(w_+)-\gamma w_+\bigr)p_-(w_+)
=0,
\end{equation}
\begin{equation}
    \bigl(v_-(w_-)-\gamma w_-\bigr)p_-(w_-)
+
\bigl(v_+(w_-)-\gamma w_-\bigr)p_+(w_-)
=0.
\end{equation}
Together with the continuity conditions at \(w_r\), these matching
conditions determine three of the four constants. The final constant is then
fixed by the normalisation condition (\ref{norm}).

\begin{figure}[t!]
    \centering
    \includegraphics[width=\linewidth]{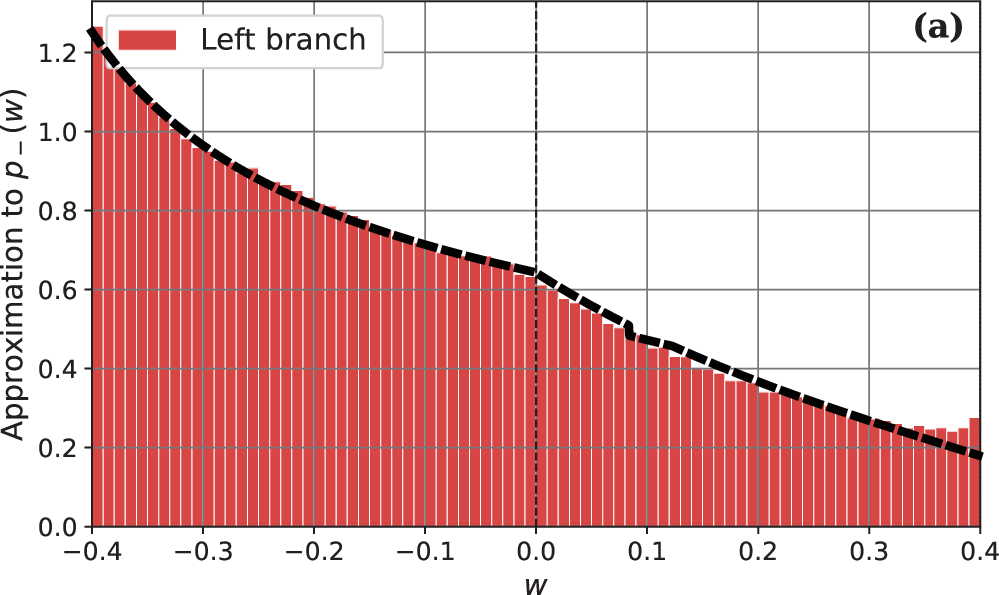}
     \includegraphics[width=\linewidth]{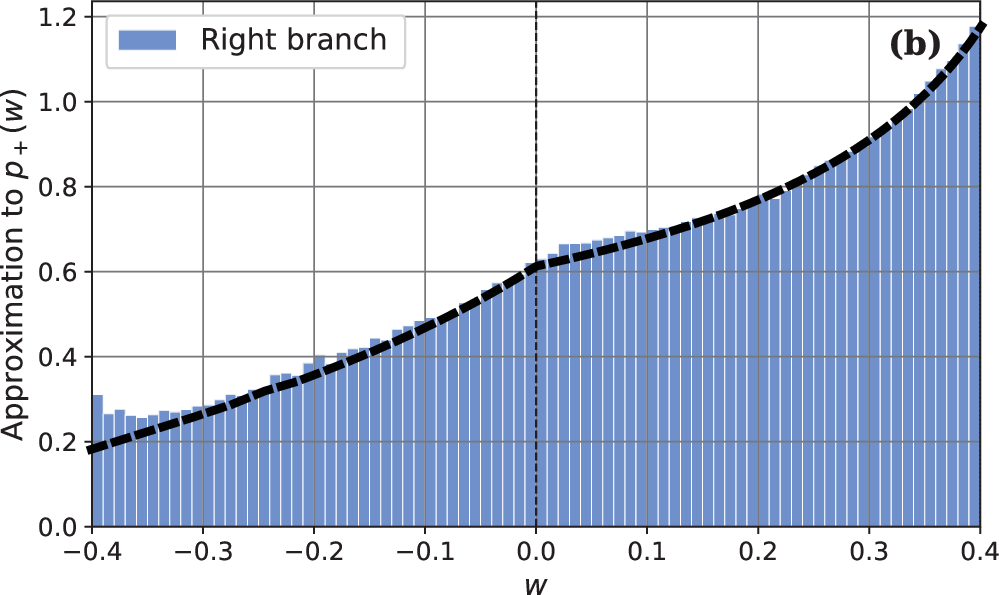}
    \caption{Numerically obtained histograms of the stationary branch densities
in the non-diffusive slow-resetting oscillatory regime with $v_r=0$
and $w_r=0$. Panel (a) shows the histogram approximation to the
left-branch density $p_-(w)$ and panel (b) shows the histogram
approximation to the right-branch density $p_+(w)$. The thick dashed curves represent the corresponding analytical density profiles shown in Fig. \ref{fig20}(a). The dashed
vertical lines mark the reset threshold. The two densities are
approximately related by reflection symmetry. Other parameters are
the same as Fig. \ref{fig20}.}
    \label{fig21}
\end{figure}

\begin{figure}[t!]
    \centering
    \includegraphics[width=\linewidth]{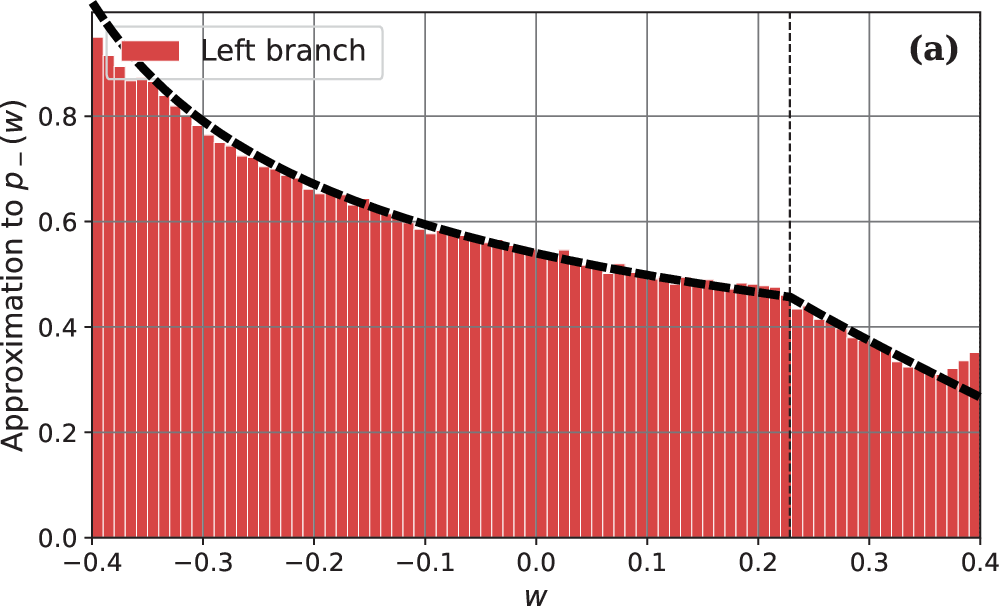}
     \includegraphics[width=\linewidth]{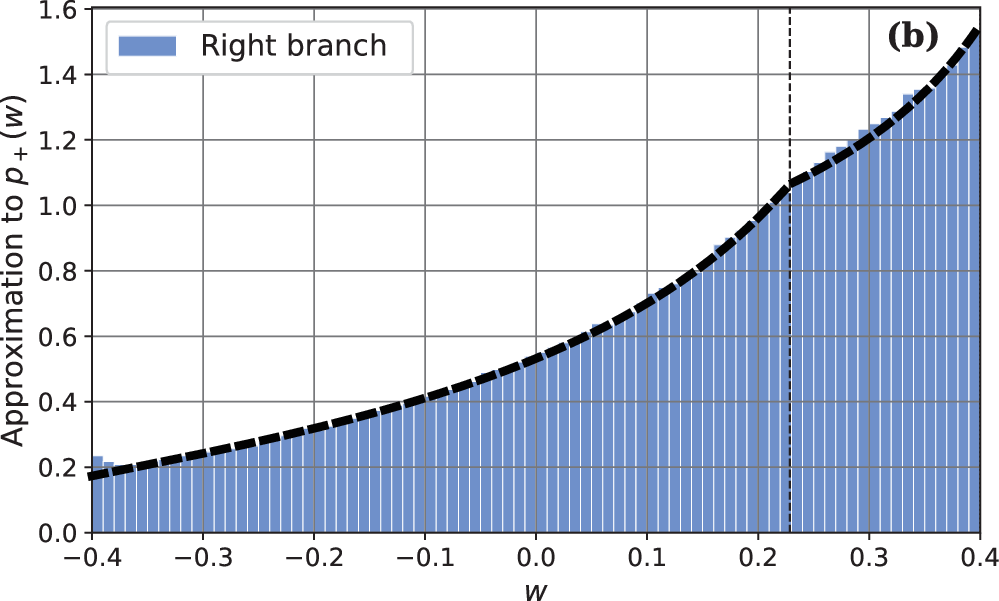}
    \caption{Same as Fig. \ref{fig20} except that
$v_r=0.2$ and $w_r\approx0.2286$. The shift in the
reset threshold breaks the symmetry between the two branch densities. The thick dashed curves represent the corresponding analytical density profiles shown in Fig. \ref{fig20}(b).}
    \label{fig22}
\end{figure}

Having determined the constants, we may now plot the resulting
stationary densities on each of the two attracting branches, as illustrated in Fig. \ref{fig20}. In the
symmetric case, where $v_r=0$, we have $w_r=0$ and the left- and
right-branch densities are reflections of one another. If instead
$v_r=0.2$, then the reset threshold shifts to
$w_r\approx0.2286$. This changes the values of $w$ for which a reset
transports the trajectory to either branch, and so the symmetry
between $p_-(w)$ and $p_+(w)$ is lost. We compare these analytical densities with direct numerical
simulations of the full slow--fast system in Figs. \ref{fig21} and \ref{fig22}. After removing an initial
transient, the values of the slow variable were separated according
to whether the trajectory was on the left or right attracting branch.
For $v_r=0$, the two empirical densities are approximately symmetric,
as expected from the analytical solution, whereas they are asymmetric when $v_r=0.2$. In both cases, the histograms reproduce the main qualitative
features of the analytical stationary densities.
 
 In the non-diffusive case, the slow dynamics is confined to the fast nullcline and resetting induces fast advection to an
outer branch determined entirely by the value of \(w\) relative to \(w_r\).
Once diffusion is incorporated, however, the slow dynamics is no longer confined to the fast nullcline so that the simple picture of a two-state switching process no longer holds. That is, between resetting events, we need to deal with the Fokker-Planck equation of the full planar dynamical system, which is non-trivial to solve.
\section{Discussion}

In this paper we explored the combined effects of stochastic resetting and diffusion applied to the fast component of the piecewise-linear McKean model. Assuming that resetting occurred on the timescale of the fast variable, we derived an averaged equation for the slow dynamics that depended on the mean of the fast variable with respect to the stationary solution of a modified second-order Fokker-Planck equation. We obtained an explicit expression for the resulting NESS in terms of parabolic cylinder functions. We then showed how the solution of the averaged equation converges to a stable fixed point that effectively determined the long-time behaviour of the full system. We thus extended our previous study of a slow-fast planar dynamical system with stochastic resetting \cite{Bressloff25} to include the effects of Gaussian white noise fluctuations in the fast variable. Although we replaced the cubic nonlinearity of the FN equations by a piecewise-linear function for the sake of analytical tractability, both models behave similarly as confirmed by numerical simulations. 

We ended the paper by considering the case of slow resetting, where
each resetting event triggers a trajectory that converges back towards the attractor of the underlying deterministic system. In the case of the McKean or FN model, the attractor is either a stable fixed point (excitable regime) or a stable limit cycle (oscillatory regime). In the particular case of the non-diffusive McKean model operating in the oscillatory regime, 
we showed that the slow variable now undergoes noisy oscillations due to resetting-induced switching between branches of the fast nullcline and derived the corresponding NESS for $w$. A major challenge for future work is understanding how to generalize the analysis of the switching system to include the effects of Gaussian fluctuations. It might be possible to make progress in the small-diffusion limit, where the slow dynamics tends to remain in a neighborhood of the fast nullcline except for rare noise-induced transitions due to resetting or large Gaussian fluctuations. 

Finally, note that in this paper we treated the non-dimensionalized McKean and FN equations as examples of a planar slow-fast dynamical system without having any specific application in mind. As we briefly mentioned in the introduction, both models can be interpreted as simplified versions of conductance-based models of a neuron with $v$ representing a  voltage and $w$ a slow recovery variable. Stimulus-induced or noise-induced transitions between the outer branches of the fast nullcline then typically signal the occurrence of an action potential. It would be interesting to explore how the results of our analysis compare to numerical studies of more realistic conductance-based models with stochastic resetting. In principle, one could also experimentally apply a resetting protocol to a real space-clamped neuron, although resetting of the voltage would no longer be instantaneous.

 \section*{Author declarations}
 \subsection*{Conflict of Interest}
 \noindent 
 The authors have no conflicts to declare
 \subsection*{Author Contributions}
\begin{small}
\noindent {\bf Paul C. Bressloff}: Conceptualization; Formal analysis. Writing - review and editing.

\noindent {\bf Jude Swaby}: Formal analysis; Numerical simulations; Writing - original draft.
\end{small}
 
 \section*{Data availability}
 
\noindent No data was generated in this study.
 
 %\bibliography{bibliography.bib}

%%%%%%%%%%%%%%%%%%%%%%%%%%%%%%%%%%%%%

%%%%%%%%%%%%%%%%%%%%%%%%%%%%%%%%%%%%%

\end{document}